\pdfoutput=1

\documentclass[12pt]{article}%
\usepackage[nosort]{cite}
\usepackage{graphicx}
\usepackage{multicol}
\usepackage{amsfonts}
\usepackage{amssymb}
\usepackage{amsmath}
\usepackage{heck}
\usepackage{setspace}
\usepackage{verbatim}
\usepackage{color}
\usepackage{epsfig}
\usepackage[margin=1in]{geometry}%
\usepackage{titletoc}%
\providecommand{\U}[1]{\protect\rule{.1in}{.1in}}
\numberwithin{equation}{section}

\newcommand{\ba}{\begin{eqnarray}}
\newcommand{\ea}{\end{eqnarray}}

\newcommand\ord{\operatorname{ord}}

\newcommand{\sutwo}{\mathfrak{su}(2)}
\newcommand{\spone}{\mathfrak{sp}(1)}

\begin{document}

\date{December 2013}

\title{On the Classification of 6D SCFTs \\[4mm] and Generalized ADE Orbifolds}

\institution{HARVARD}{\centerline{${}^{1}$Jefferson Physical Laboratory, Harvard University, Cambridge, MA 02138, USA}}

\institution{UCSBmath}{\centerline{${}^{2}$Department of Mathematics, University of California Santa Barbara, CA 93106, USA}}

\institution{UCSBphys}{\centerline{${}^{3}$Department of Physics, University of California Santa Barbara, CA 93106, USA}}

\authors{Jonathan J. Heckman\worksat{\HARVARD}\footnote{e-mail: {\tt jheckman@physics.harvard.edu}},
David R. Morrison\worksat{\UCSBmath, \UCSBphys}\footnote{e-mail: {\tt drm@physics.ucsb.edu}}, and
Cumrun Vafa\worksat{\HARVARD}\footnote{e-mail: {\tt vafa@physics.harvard.edu}}}

\abstract{We study $(1,0)$ and $(2,0)$ 6D superconformal field theories (SCFTs) that can be constructed in F-theory.
Quite surprisingly, all of them involve an orbifold singularity $\mathbb{C}^{2} / \Gamma$ with $\Gamma$ a
discrete subgroup of $U(2)$. When $\Gamma$ is a subgroup of $SU(2)$, all discrete subgroups are allowed, and
this leads to the familiar ADE classification of $(2,0)$ SCFTs.  For more general
$U(2)$ subgroups, the allowed possibilities for $\Gamma$ are not arbitrary
and are given by certain generalizations of the A- and D-series.  These
theories should be viewed as the minimal 6D SCFTs.  We obtain all other SCFTs by
bringing in a number of E-string theories and/or decorating curves in the base by non-minimal gauge algebras.
In this way we obtain a vast number of new 6D SCFTs, and we conjecture that our
construction provides a full list.}

\maketitle

\tableofcontents

\enlargethispage{\baselineskip}

\setcounter{tocdepth}{2}

\newpage

\section{Introduction \label{sec:INTRO}}

A striking prediction of string compactification is the existence of
interacting conformal fixed points in six dimensions. Such theories resist a
UV Lagrangian description and involve tensionless strings in the low energy theory.

Theories with $(2,0)$ supersymmetry in six dimensions admit a simple ADE
classification. This is transparently realized by IIB string theory on the
background $\mathbb{R}^{5,1}\times\mathbb{C}^{2}/\Gamma$, with $\Gamma$ a
discrete ADE subgroup of $SU(2)$ \cite{Witten:1995zh}. In this description,
the tensionless strings of the theory are realized by D3-branes wrapped on
collapsing two-cycles of the internal geometry. For the A-series, the dual
M-theory description is a stack of coincident M5-branes in flat space
\cite{Strominger:1995ac}.

Comparatively far less is known about six-dimensional $(1,0)$ theories. The
most well-known example of this type is the E-string theory, which can be
defined as the theory of a small $E_{8}$ instanton in heterotic $E_{8}\times
E_{8}$ string theory \cite{WittenSmall, Ganor:1996mu, Seiberg:1996vs}. In this
description, the $E_{8}$ gauge symmetry of the heterotic theory becomes a
flavor symmetry of the superconformal theory.
Upon further compactification on a $T^{2}$, this provides a
realization of the 4D $\mathcal{N}=2$ Minahan-Nemeschansky theory with $E_{8}$
flavor symmetry \cite{MNI, MNII}. Additional examples of $(1,0)$ theories
include the collision of E-type singularities in compactifications of F-theory
\cite{Bershadsky:1996nu}, five-branes placed at ADE singularities
\cite{Blum:1997mm}, and many others
\cite{Witten:1996qz,
Seiberg:1996qx,
Aspinwall:1996vc,
Intriligator:1997kq,
BershadskyVafaAnomalies,
Blum:1997fw,
Aspinwall:1997ye,
Seiberg:1997zk,
Intriligator:1997dh}.
However, a systematic classification of $(1,0)$ theories
has not been undertaken.

In this note we classify all possible six-dimensional SCFTs which arise from
singular limits of an F-theory compactification. This includes all the
previously known examples that we are aware of as well as several new classes of
theories. We are therefore led to conjecture that we have the full
list.

The basic idea will be to consider F-theory on an elliptically fibered Calabi-Yau
threefold $\pi:X\rightarrow B$ over a non-compact complex surface $B$. Gravity also decouples
in such cases. When compact two-cycles of $B$ collapse to zero size, we expect to reach a
conformal fixed point with $(1,0)$ or $(2,0)$ supersymmetry, signaled by the
presence of tensionless strings coming from D3-branes wrapped over the
two-cycles. Roughly speaking, the two-cycles of the base
$B$ should be viewed as contributing tensionless strings to the 6D theory,
while the two-cycles of the fiber contribute massless particles. The
combination of the two produces a rich class of possible six-dimensional
superconformal field theories.

Some of the previously encountered 6D SCFTs have a simple realization in F-theory. For example,
the $(2,0)$ theories correspond to the special case where the F-theory geometry is $B \times T^2$ with
$B = \mathbb{C}^{2}/ \Gamma$ for $\Gamma$ a discrete subgroup of $SU(2)$. Resolving the base gives
a configuration of $-2$ curves intersecting according to an ADE Dyknin diagram. As another example,
the E-string theory corresponds to the generic elliptic Calabi-Yau where the base $B$
is the total space $\mathcal{O}(-1) \rightarrow\mathbb{P}^{1}$
\cite{Witten:1996qb,MorrisonVafaII}. The conformal fixed point corresponds to
shrinking the base $\mathbb{P}^1$ to zero size.

An especially important class of six-dimensional SCFTs that we discover are the ``minimal'' ones
with generic elliptic fibration and with no superfluous E-string theories. Quite surprisingly,
this class of SCFTs are all defined by F-theory with base given by an orbifold
$\mathbb{C}^{2}/\Gamma$ for $\Gamma$ some discrete subgroup of $U(2)$. Moreover,
not all subgroups $\Gamma$ are allowed. We classify all discrete subgroups
consistent with the existence of an elliptic fibration.

To reach this result, we proceed by a process of elimination. Using the
classification results on ``non-Higgsable'' six-dimensional F-theory models
\cite{Morrison:2012np} and by determining how these non-Higgsable clusters can
couple to one another (i.e., their compatibility with the elliptic fibration) we
find that the configuration of such curves reduces to an ADE graph consisting
of just curves with self-intersection $-2$, or a generalized A- or D-type graph of curves where
the self-intersection can be less than $-2$. Moreover,
we show that all such graphs arise as a collection of intersecting $\mathbb{P}%
^{1}$'s which blow down to orbifolds of $\mathbb{C}^{2}$, by specific discrete subgroups
of $U(2)$. Scanning through all possibilities, we determine the full list of
such orbifold singularities.

For the generalized A-type and D-type theories, we find the following infinite
families of ``rigid'' minimal theories:
\begin{align}
\mathcal{C}_{\text{rigid}}^{(A)} &  =\alpha A_{N}\beta\text{ \ \ for
\ \ }N\geq0\\
\mathcal{C}_{\text{rigid}}^{(D)} &  =D_{N}\gamma\text{ \ \ \ \ for \ \ }N\geq2
\end{align}
where $A_{N}$ and $D_{N}$ refer to a Dynkin diagram with $2$ at each node, and the linear part of the $D_N$ series attaches to $\gamma$. Here,
$\alpha$, $\beta$ and $\gamma$ are generalized A-type subgraphs of up to five nodes:
\begin{align}
\alpha &  \in\left\{  7,33,24,223,2223,22223\right\}  \\
\beta &  \in\left\{  7,33,42,322,3222,32222\right\}  \\
\gamma &  \in\left\{  24,32\right\}  ,
\end{align}
where each label denotes the self-intersection with signs reversed. The rigid theories
have the property that increasing the integer at any node would lead to an inconsistent F-theory model. Moreover,
by lowering the label to an integer which is at least $2$, we reach all the other infinite families of minimal $(1,0)$ SCFTs.
In addition to these infinite families, we also determine all outliers. We find that these only occur in the
A-type series for nine or fewer nodes.

Generically, a minimal 6D SCFT can be used to construct
non-minimal SCFTs by bringing in additional ingredients: We can bring
in a number of E-string theories and/or make the
Kodaira-Tate type of the fiber over the curves more singular. This amounts to decorating curves
in the base with a self-consistent choice of gauge algebra. Activating a deformation of the
gauge algebra then generates a flow to a lower theory. This process
can be done for the $(2,0)$ ADE theories, as well as for many of the other minimal 6D SCFTs.
In fact in this way we obtain all the known 6D SCFTs that we are aware of
and we conjecture that this construction leads to the full list.

A striking outcome of our analysis is the appearance of discrete subgroups of
$U(2)$. Strings on orbifolds of $\mathbb{C}^{2}$ which are not strictly
contained in $SU(2)$ would lead to a non-supersymmetric compactification, with
a closed string tachyon. For work in this direction, see for example
\cite{Adams:2001sv, Vafa:2001ra, David:2001vm, Martinec:2002wg,c3zn,tachyons}.
The reason we reach a supersymmetric vacuum in our case is that the full
F-theory model is still a Calabi-Yau threefold, which generically happens by
making the string coupling constant position dependent and of order one. In
some very special cases, this arises by embedding the $U(2)\subset SU(3)$ and
having the group $\Gamma$ also act on the elliptic fiber to preserve
supersymmetry \cite{Witten:1996qb}.

The rest of this paper is organized as follows. First, in section \ref{sec:6D}
we review some relevant aspects of F-theory in six dimensions, and in
particular the structure of \textquotedblleft non-Higgsable
clusters\textquotedblright\ which appears in the classification of
six-dimensional supergravity theories. In section \ref{sec:SCFT} we turn our
attention to six-dimensional SCFTs generated by singular limits of F-theory
compactifications. Section \ref{sec:ENDPOINT} is the first step in our
classification program, where we show that all minimal 6D SCFTs are
obtained from ADE type graphs of curves. In section \ref{sec:LIST} we give the full list of
minimal 6D SCFTs. In section \ref{sec:SUPP} we explain in broad terms
how to supplement these minimal theories by additional blowups in the base, and
more singular behavior in the elliptic fibration. In section \ref{sec:FLOWS} we make some preliminary
comments on the interpretation of blowdowns as triggering RG flows. We
present our conclusions in section \ref{sec:CONC}. Some additional
mathematical material is collected in a set of Appendices. We also include
a brief set of instructions on usage of the \texttt{Mathematica} notebook included
with the \texttt{arXiv} submission.

\section{6D\ Theories From F-theory \label{sec:6D}}

To frame our discussion, in this section we review some aspects of F-theory in
six dimensions. Since we are dealing with theories with six flat spacetime
dimensions and eight real supercharges, we shall be concerned with F-theory
compactified on $\widetilde{X}$, a smooth elliptically fibered Calabi-Yau threefold
$\widetilde{\pi}:\widetilde{X}\rightarrow B$ with a section, where $B$ is a complex
surface and all of the fibers of $\widetilde{\pi}$ have dimension one.\footnote{If
the condition on the fibers is not satisfied, it is possible to blow up the
base and possibly perform some flops which gives a new Calabi-Yau
threefold satisfying the condition \cite{Grassi91}.}

Blowing down all fiber components which do not meet the section leads to
a singular fibration $\pi:X\rightarrow B$ which always has an equation
in Weierstrass form:%
\begin{equation}
y^{2}=x^{3}+fx+g , \label{Weierstrass}%
\end{equation}
where $f$ and $g$ are sections of the line bundles
$\mathcal{O}(-4K_{B})$ and $\mathcal{O}%
(-6K_{B})$, respectively. The discriminant locus, which marks where the
elliptic fibers are singular, is a section of
$\mathcal{O}(-12K_{B})$ given by:%
\begin{equation}
\Delta=4f^{3}+27g^{2},
\end{equation}
and should be viewed as indicating the location of seven-branes in the system.
If we can reverse the blowdown process, i.e., if the singularities in
equation (\ref{Weierstrass}) can be resolved to a Calabi-Yau threefold
$\widetilde{X}$ with all fibers one-dimensional, then $X$ determines a
nonsingular F-theory model, which is a
six-dimensional theory with some number of
normalizable tensor multiplets, vector multiplets, and hypermultiplets. Our
primary interest is in the class of six-dimensional superconformal field
theories we can achieve in these models. For this reason, we shall focus on
the case where $B$ is non-compact. This means that a number of the tight
restrictions on six-dimensional supergravity theories will not apply.
Indeed, this is reflected in the comparative ease with which we
will be able to generate various elliptic models.

Each component of the discriminant locus
is associated with a seven-brane (or stack of seven-branes)
wrapping a curve $\Sigma\subset B$. Each
such seven-brane supports a gauge algebra\footnote{The global structure of
the gauge group is more subtle \cite{Aspinwall:1998xj}.} $\mathfrak{g}_{\Sigma}$ which
is dictated by the order of vanishing of $f$, $g$ and $\Delta$ along $\Sigma$.
To obtain a nonsingular F-theory model, along each curve $\Sigma$
we must either have $f$ vanishing to
order less than $4$, or $g$ vanishing to order less than $6$.\footnote{This
condition is often stated in a way that includes $\Delta$ vanishing
to order less than $12$, but since $\Delta=4f^3+27g^2$, if $f$ vanishes
to order at least $4$ and $g$ vanishes to order at least $6$, it will
automatically be true that $\Delta$ vanishes to order at least $12$.}
Higher order vanishing leads to a model with the unfortunate property
that on any blowup of $X$ to a smooth space $\widetilde{X}$, the three-form
on $\widetilde{X}$ always has poles.
 This strongly suggests such models are
unphysical, and they will not be considered here.

The singularities of the elliptic fibration can become worse at the intersection
of two components of the discriminant locus; such points are associated with matter
fields trapped at the intersection of two seven-branes. An important subtlety
here is that to fully specify the field content of the theory, one must start
with the smooth Calabi-Yau $\widetilde{X}$ and only then take a degeneration
limit.\footnote{For a recent alternative approach which avoids the need to
resolve $X$, see \cite{Grassi:2013kha}}

In the maximally Higgsed phase, i.e., after activating all possible
hypermultiplet vevs, it is possible to classify the resulting models in terms
of the geometry of the base $B$ \cite{Morrison:2012np}.\footnote{Note
that the analysis of \cite{Morrison:2012np} was carried out in the context
of a compact base, but the same analysis applies to the case of
a neighborhood of a collection of compact curves which we study here.}
In the base geometry,
there is a configuration of curves which support seven-brane gauge theories.
Such configurations of curves can be decomposed into a set of
\textquotedblleft non-Higgsable clusters\textquotedblright\ which only intersect
curves with self-intersection $-1$.

Let us briefly review this classification result \cite{Morrison:2012np}. For a
seven-brane wrapping a compact curve $\Sigma$, we can analyze the geometry
normal to $\Sigma$ inside of $B$ via its self-intersection. In the maximally
Higgsed phase, all possible hypermultiplet vevs are activated. This means, for
example, that non-abelian gauge symmetries can only be supported on rigid
curves, i.e., ones where the seven-brane cannot move around. Such curves are
$\mathbb{P}^{1}$'s, with self-intersection $-n<0$ for some natural number $n$,
and the local geometry is the total space $\mathcal{O}(-n)\rightarrow
\mathbb{P}^{1}$. When $n=1$ and $n=2$, one can define a smooth Weierstrass
model, i.e., no gauge symmetry is present in the maximally Higgsed phase.

When $n>2$, there is a gauge symmetry, and the \textit{minimal} order of
vanishing of $f$, $g$ and $\Delta$ along $\Sigma$ is dictated by $n$. The full
list of isolated curves which support a gauge symmetry are theories with an
$\mathfrak{e}_{8}$ seven-brane gauge theory coupled to some number of small
instantons, or a seven-brane possibly coupled to a half hypermultiplet. The
full list of possibilities is:%
\begin{equation}
\begin{tabular}
[c]{|l|l|l|l|l|l|l|l|}\hline
7-brane
$\backslash$
cluster & $3$ & $4$ & $5$ & $6$ & $7$ & $8$ &$12$\\\hline
$\mathfrak{g}_{\Sigma}$ & $\mathfrak{su}(3)$ & $\mathfrak{so}(8)$ &
$\mathfrak{f}_{4}$ & $\mathfrak{e}_{6}$ & $\mathfrak{e}_{7}$ & $\mathfrak{e}%
_{7}$ & $\mathfrak{e}_8$ \\\hline
Hyper & none & none & none & none & $\frac{1}{2}\mathbf{56}$ & none & none\\\hline
\end{tabular}
\end{equation}
where in the second row  we have indicated the gauge
algebra, and in the third row  we have indicated the
hypermultiplet content. If we further tune $f$ and $g$, in general we may
obtain higher gauge groups and other matter content which can be Higgsed down
to the minimal form above.

For curves of self-intersection $-9$, $-10$, and $-11$, we again find
gauge symmetry $\mathfrak{e}_8$.  However, in these cases there are
``small instantons'' as well.
For example, in the $n=9$ theory, this is associated
with a minimal Weierstrass model:%
\begin{equation}
y^{2}=x^{3}+g_{3}(z^{\prime})z^{5},
\end{equation}
where $z$ and $z^{\prime}$ are respectively affine coordinates normal to, and
along the curve $\Sigma$ so that $z=0$ is the location of the curve $\Sigma$,
and $g_{3}(z^{\prime})$ is a degree three polynomial, and its zeroes specify
the locations of \textquotedblleft small instantons\textquotedblright. This
terminology is borrowed from the heterotic dual description, where it was
shown that the $\mathbb{P}^{1}$ with self-intersection $-n$ is dual to a strong
coupling point of heterotic string theory for maximally Higgsed $E_{8}$ with
instanton number $12-n$ \cite{MorrisonVafaII,Seiberg:1996vs}.
The presence of small instantons means that we do not
have a nonsingular F-theory model, but this is repaired by blowing up the
base at the location of the zeros of $g_{3}(z^{\prime})$, yielding
a curve of self-intersection $-12$  meeting $12-n$  curves of
self-intersection\footnote{There are other possibilities if
the zeroes of $g_{3}(z^{\prime})$ have non-trivial multiplicity.} $-1$.
When the base is blown up in this way, there is an appropriate
modification\footnote{Since the canonical bundle of the blownup base
gains an order of vanishing along the $-1$ curve, the orders of vanishing
of $f$ and $g$ along that curve must be reduced by $4$ and $6$, respectively,
by dividing by a power of the local equation of the curve.}
of the Weierstrass equation which produces a potentially
nonsingular F-theory model.
This is our first encounter
with the primary issue we must deal with in this paper:  determining
how much to blow up a given base in order to produce a nonsingular
F-theory model.

In addition to isolated curves, there are a few non-Higgsable clusters
which have multiple $\mathbb{P}^{1}$'s. These configurations are of the form
$n_{1},...,n_{r}$, corresponding to $\mathbb{P}^{1}$'s with self-intersection
$-n_{i}$ which intersect pairwise at a single point dictated by their location
on the string of integers. As found in \cite{Morrison:2012np}, the full list
of such configurations is:%
\begin{equation}%
\begin{tabular}
[c]{|l|l|l|l|}\hline
7-brane
$\backslash$
cluster & $3,2$ & $3,2,2$ & $2,3,2$\\\hline
$\mathfrak{g}_{\Sigma}$ & $\mathfrak{g}_{2}\oplus\sutwo$ &
$\mathfrak{g}_{2}\oplus\spone$ & $\sutwo\oplus
\mathfrak{so}(7)\oplus\sutwo$\\\hline
Hyper & $\frac{1}{2}\left(  \mathbf{7}+\mathbf{1},\mathbf{2}\right)  $ &
$\frac{1}{2}\left(  \mathbf{7}+\mathbf{1},\mathbf{2}\right)  $ & $\frac{1}%
{2}\left(  \mathbf{2},\mathbf{7}+\mathbf{1},\mathbf{1}\right)  \oplus\frac
{1}{2}\left(  \mathbf{1},\mathbf{7}+\mathbf{1},\mathbf{2}\right)  $\\\hline
\end{tabular}
\ \ .
\end{equation}
An interesting feature of these clusters is that they contain curves of
self-intersection $-2$, which support a gauge group factor. This is because
these curves intersect another curve which already has a singular fiber, which
is then inflicted on its neighbor. Observe that in the $3,2,2$ configuration,
the $-3$ curve and
the middle $-2$ curve carry gauge symmetry algebras, but the $-2$ curve
at the end does not.  We label the symmetry algebra for the middle $-2$
curve as $\spone$ rather than $\sutwo$ because it arises in a different
way in F-theory.  For additional details, see Appendix \ref{APP:gluing}.

Finally, in addition to the non-Higgsable clusters which carry a gauge
symmetry, we can also have isolated configurations without a gauge group.
These are given by $-2$ curves which intersect according to an ADE Dynkin diagram.

Now, given a collection of such non-Higgsable clusters, we get a local model
for an F-theory base by \textquotedblleft gluing\textquotedblright\ such
configurations together using curves of self-intersection $-1$. Labelling two such configurations of curves as
$\mathcal{C}_{(1)}$ and $\mathcal{C}_{(2)}$, we get a new configuration of
curves $\mathcal{C}_{\text{new}}$ by introducing an additional $-1$ curve:%
\begin{equation}
\mathcal{C}_{\text{new}}=(\mathcal{C}_{(1)},1,\mathcal{C}_{(2)}),
\end{equation}
i.e., we have a $-1$ curve which intersects a curve $\Sigma_{(1)}$ appearing
in the configuration $\mathcal{C}_{(1)}$, and a curve $\Sigma_{(2)}$ appearing
in the configuration $\mathcal{C}_{(2)}$.

A priori, there are many ways to piece together such non-Higgsable clusters to
form much larger configurations. However, at each stage of gluing, there is a consistency
condition which must be satisfied. The two
curves $\Sigma_{(1)}$ and $\Sigma_{(2)}$ which we glue via a $-1$ curve
must have appropriate order of vanishing along $f$, $g$ and $\Delta$ to be
consistent with having a smooth phase (after resolution). Algebraically, this requires the combined gauge
algebra $\mathfrak{g}_{(1)}\oplus\mathfrak{g}_{(2)}\subset\mathfrak{e}_{8}$
(as we verify in Appendix~\ref{APP:gluing}).
An interesting feature of this condition is that a $-6$ curve can connect to a
$-3$ curve, provided the $-3$ curve supports a gauge algebra $\mathfrak{su}%
(3)$. If, however, the $-3$ curve supports a $\mathfrak{g}_{2}$ gauge algebra
(as happens for some NHCs), then the self-intersection must be $-5$ or more
since $\mathfrak{f}_{4}\oplus\mathfrak{g}_{2}$ is a subalgebra of
$\mathfrak{e}_{8}$, but $\mathfrak{e}_{6}\oplus\mathfrak{g}_{2}$ is not a
subalgebra of $\mathfrak{e}_{8}$.

Finally, if the elliptic fibration is trivial, and only then, it is also
possible to add probe five-branes to an F-theory model (otherwise the identity
of the five-brane is ambiguous due to $SL(2,\mathbb{Z})$ monodromy), and in
addition including the action of orientifolds. This in particular can only be
done in the case where we simply have an ADE singularity, and add a number $k$
of D5-branes, with or without orientifold 5-planes, as was studied in
\cite{Blum:1997mm}. However, as we explain in section \ref{sec:SUPP},
this can also be constructed without using five-branes by considering
ADE singularities with additional restrictions on the
Kodaira-Tate fiber type over the singularity.

\section{Building Blocks of 6D SCFTs \label{sec:SCFT}}

Our interest in this paper concerns the six-dimensional superconformal field
theories (SCFTs) generated by singular limits of an F-theory compactification.
To reach such SCFTs, we need to take a limit where two-cycles of the base $B$
collapse to zero size. Clearly, the choice of SCFT is dictated by a
configuration of smooth curves $\mathcal{C}$, which we blow down to reach a
singular configuration $\mathcal{C}_{\text{sing}}$.

This amounts to a \textquotedblleft global\textquotedblright\ requirement that we can
simultaneously contract all of the curves $\Sigma_i$ of negative self-intersection to
zero. Introducing the adjacency matrix:\footnote{It is more common to use
a central dot to denote the intersection number, which we shall do for the rest of this paper;
when the curves are distinct, this means the number
of intersection points counted with multiplicity.}%
\begin{equation}
A_{ij}=-(\Sigma_{i}\cap\Sigma_{j}),
\end{equation}
this requires $A_{ij}$ to be positive definite, i.e., all eigenvalues must be
positive \cite{MR0153682,MR0146182,MR0137127}.
In general, this is a non-trivial condition which excludes many
candidate configurations. As an example of an inadmissible configuration, we
cannot have two $-1$ curves intersect, because the adjacency matrix is not
positive definite. A related comment is that in any configuration of
non-Higgsable clusters, we get a tree-like structure with no closed loops.
As explained in Appendix~\ref{APP:constraints}, there cannot be a physical F-theory
model with a loop of contractible curves in the base, for in that
case $f$ and $g$
 would  vanish to orders at least $4$ and $6$, respectively, along each
 of the curves in the loop.

From this perspective, the geometric objects of interest are all possible ways
that we can arrive at a singular base $B$ which admits \textit{some}
resolution to a smooth F-theory model. These different choices of resolution
then parameterize different phases of the theory where all states have picked
up some tension or mass.\footnote{In six dimensions, the blowups of the fiber
are not part of the Coulomb branch. They are, however, part of the Coulomb
branch for the five-dimensional theories defined by a further compactification
on a circle.} Our task is therefore to classify all the physically distinct
SCFTs which arise for some choice of $\mathcal{C}_{\text{sing}}$, and to then determine all possible
elliptic fibrations over a given base.

Clearly, there are many choices of non-singular $B$, so even specifying the full
list of bases would at first appear to be a daunting task. As a warmup, in this section
we shall therefore discuss some examples of the types of behavior we can
expect to encounter. First, we discuss how the $(2,0)$ theories come about in
F-theory. Then, we discuss the case of the E-string theories, which serve as
the \textquotedblleft glue\textquotedblright\ joining various non-Higgsable
clusters. After this, we turn to the theories defined by a single
non-Higgsable cluster. Later, we perform a classification of all possible 6D
SCFTs which can be realized by putting these ingredients together.

\subsection{$(2,0)$ Theories}

Let us begin by showing how all the $(2,0)$ theories come about in our
classification. These are specified by a configuration of $-2$ curves
intersecting according to the appropriate ADE\ Dynkin diagram. The full list
consists of two infinite series:%
\begin{equation}
A_{N}:\underset{N}{\underbrace{%
\begin{tabular}
[c]{|l|l|l|}\hline
$2$ & $...$ & $2$\\\hline
\end{tabular}
\ }}\text{, \ \ }D_{N}:\underset{N-1}{\underbrace{%
\begin{tabular}
[c]{|l|l|l|l|}\hline
& $2$ &  & \\\hline
$2$ & $2$ & $...$ & $2$\\\hline
\end{tabular}
\ }}%
\end{equation}
and the exceptional configurations:%
\begin{equation}
E_{6}:%
\begin{tabular}
[c]{|l|l|l|l|l|}\hline
&  & $2$ &  & \\\hline
$2$ & $2$ & $2$ & $2$ & $2$\\\hline
\end{tabular}
\ \text{, \ \ }E_{7}:%
\begin{tabular}
[c]{|l|l|l|l|l|l|}\hline
&  & $2$ &  &  & \\\hline
$2$ & $2$ & $2$ & $2$ & $2$ & $2$\\\hline
\end{tabular}
\ \text{, \ \ }E_{8}:%
\begin{tabular}
[c]{|l|l|l|l|l|l|l|}\hline
&  & $2$ &  &  &  & \\\hline
$2$ & $2$ & $2$ & $2$ & $2$ & $2$ & $2$\\\hline
\end{tabular}
\ \ \ \ .
\end{equation}
The special feature of all of these examples is that the base $B$ defines a
non-compact Calabi-Yau, i.e., it can be represented as an orbifold
$\mathbb{C}^{2}/\Gamma$ for $\Gamma$ discrete subgroup of $SU(2)$. Since the
base $B$ is already Calabi-Yau, the elliptic fibration of the model can be
trivial, i.e., we can take $B\times T^{2}$. This means we get theories
with $(2,0)$ supersymmetry. In fact, the ADE\ classification of such discrete
subgroups ensures that this is the full list of $(2,0)$ theories.

\subsection{E-String Theory}

Most of the 6D theories we consider will only have eight real supercharges.
This means the complex base $B$ will not be Calabi-Yau, and the only way to
satisfy the conditions of supersymmetry will be via a non-trivial elliptic
fibration. Perhaps the simplest example of this type is the E-string theory
defined by the base $\mathcal{O}(-1)\rightarrow\mathbb{P}^{1}$. We reach the
conformal phase by shrinking the $\mathbb{P}^{1}$ to zero size.

In some sense, this theory is the \textquotedblleft glue\textquotedblright%
\ which holds together all of the other $(1,0)$ theories, so it is worthwhile
to discuss it in more detail. One way to arrive at the E-string is to start
with the minimal Weierstrass model \cite{MorrisonVafaI, MorrisonVafaII} (see
also \cite{WittenSmall, Ganor:1996mu, Seiberg:1996vs, Witten:1996qb}):
\begin{equation}
y^{2}=x^{3}+z^{\prime}z^{5}, \label{minusOne}%
\end{equation}
where $(z,z^{\prime})$ are coordinates on $\mathbb{C}^{2}$. There is a
singularity at the intersection $z^{\prime}=z=0$, which is the location of the
small instanton. Blowing up the intersection point, we introduce an additional
$\mathbb{P}^{1}$, with local geometry $\mathcal{O}(-1)\rightarrow
\mathbb{P}^{1}$. In this geometry, there is a non-compact seven-brane along
the locus $z=0$, corresponding to an $E_{8}$ flavor symmetry. Returning to our discussion of gluing
NHCs by $-1$ curves given in section \ref{sec:6D} and Appendix \ref{APP:gluing}, observe that
we can interpret this gluing construction as gauging a product subalgebra of the $\mathfrak{e}_{8}$
flavor symmetry.

There is also a heterotic M-theory dual description of the E-string theory.
Starting with a single $E_{8}$ wall in flat space, we can introduce an
additional M5-brane to probe this theory. The distance of the M5-brane from
the wall corresponds to the overall size of the $\mathbb{P}^{1}$. In the limit
where the $\mathbb{P}^{1}$ shrinks to zero size, the M5-brane touches the
wall, and is better viewed as a singular gauge field configuration
corresponding to a pointlike instanton.

Now, equation (\ref{minusOne}) is but one presentation of this singular
geometry. We can maintain the same singular behavior by including all terms
with the same degree along the locus $z=z^{\prime}$. In the Weierstrass model,
this means we can also consider the presentations:%
\begin{equation}
y^{2}=x^{3}+f_{4}(z,z^{\prime})x+g_{6}(z,z^{\prime}),
\end{equation}
with $f_{4}$ and $g_{6}$ homogeneous polynomials of respective degrees $4$ and
$6$ in the variables $z$ and $z^{\prime}$. As this example makes clear, for
some presentations the overall flavor symmetry
can be obscured by the choice of equation. However, all of
these different choices amount to irrelevant
deformations of the 6D theory.

\subsection{Single Clusters and Orbifolds \label{ssec:CLUSTERcft}}

The non-Higgsable clusters themselves yield a rich class of $(1,0)$ theories. Recall
that the non-Higgsable clusters contain some configuration of isolated
curves which are simultaneously contractible. All of
these examples are specified by a linear chain of $\mathbb{P}^{1}$'s with
negative self-intersection. Though not given this name, some examples of
this type were considered in \cite{Witten:1996qb}.

Upon blowdown, all of these configurations define an orbifold of
$\mathbb{C}^{2}$ by a discrete cyclic subgroup of $U(2)$. Indeed, a well-known
result from the theory of Hirzebruch--Jung resolutions
\cite{jung,MR0062842,Riemen:dvq} is that a configuration of curves:
\begin{equation}
\mathcal{C}=x_{1}\cdots x_{r}%
\end{equation}
under blowdown defines the orbifold $\mathbb{C}^{2}/\Gamma$, with $\Gamma$ a
discrete subgroup of $U(2)$ generated by the group action:%
\begin{equation}
(z_{1},z_{2})\mapsto(\omega z_{1},\omega^{q}z_{2})\text{ \ \ where
\ \ }\omega=e^{2\pi i/p}\text{
\ \ and \ \ }\frac{p}{q}=x_{1}-\frac{1}{x_{2}-...\frac{1}{x_{r}}}.
\end{equation}

The specific orbifold for all of the isolated $-n$ theories is:%
\begin{equation}
-n\text{ theories: \ \ }(z_{1},z_{2})\mapsto(\omega z_{1},\omega z_{2})\text{
\ \ for \ \ }\omega=\exp(2\pi i/n),
\end{equation}
while for the clusters with more than one curve, we instead have:%
\begin{equation}%
\begin{tabular}
[c]{|l|l|l|l|}\hline
cluster: & $3,2$ & $3,2,2$ & $2,3,2$\\\hline
$p/q$ & $5/2$ & $7/3\text{ }$ & $8/5$\\\hline
\end{tabular}
\ \ \ .
\end{equation}
Note that reading the configuration from right to left determines
a second generator, which is $5/3$ for $3,2$ and $7/5$ for $3,2,2$.

An important feature of these orbifolds is that they are only supersymmetric
in the context of F-theory, and not in the more limited context of
perturbative IIB\ string theory. This is because the orbifolds are specified
by a $U(2)$ group action which does not embed in $SU(2)$. Nevertheless,
supersymmetry is preserved because the profile of the axio-dilaton is non-trivial.

For most of the $-n$ curve theories, the orbifold group action also extends to the elliptic fibration, and we
can write the F-theory model as $(\mathbb{C}^2 \times T^2) / \mathbb{Z}_{n}$. Letting $\lambda = dx / y$ denote the holomorphic
differential of the constant $T^2$, to get a Calabi-Yau threefold,
the holomorphic three-form $dz_1 \wedge dz_2 \wedge \lambda$ must be invariant under the group action, so we have
$(z_1 , z_2 , \lambda) \mapsto (\omega z_1 , \omega z_2 , \omega^{-2} \lambda)$. For the orbifold
to act on the $T^2$ directions, we also need $\omega^2$ to be of order $1,2,3,4,6$, which in turn means that such a
presentation is available for $n=2,3,4,6,8,12$. Observe, however, that the NHCs with $n=5$ and $n = 7$ are
not of this type. For further discussion, see \cite{Witten:1996qb}.

As a final remark, let us note that the isolated $-n$ curve theories can all
be realized in a decoupling limit of F-theory on the Hirzebruch surface
$\mathbb{F}_{n}$. These complex surfaces are given by a $\mathbb{P}^{1}$
bundle over a base $\mathbb{P}^{1}$ where the self-intersection of the base is
$-n$. For further details of F-theory on Hirzebruch surfaces, see e.g.
\cite{Witten:1996qb, MorrisonVafaI, MorrisonVafaII, BershadskyPLUS}.

\subsection{A Plethora of Bases}

Clearly, there are a vast number of possible F-theory bases, corresponding to a choice of
tree-like structure. For example, it is possible to generate
several infinite families of SCFTs by stringing together NHCs into arbitrarily long chains. This
includes, for example, some of the chains of NHCs discussed in section 4.3 in \cite{Morrison:2012np}.

A priori, however, given a chain of NHCs, there is no guarantee that it is possible to simultaneously
contract all of the curves in the base. Though rather inefficient, this can be checked on a case by case basis.
To illustrate some of these issues, consider the configuration of curves:
\begin{equation}
\mathcal{C}=313\text{.}%
\end{equation}
This satisfies the local condition of non-Higgsable clusters, and as can be
verified, the adjacency matrix:%
\begin{equation}
A_{\mathcal{C}}=\left[
\begin{array}
[c]{ccc}%
3 & -1 & 0\\
-1 & 1 & -1\\
0 & -1 & 3
\end{array}
\right]  \text{,}%
\end{equation}
is positive definite. However, the seemingly rather similar configuration:%
\begin{equation}
\mathcal{C}=13131
\end{equation}
does not have a positive definite adjacency matrix.
Rather than directly computing the eigenvalues, a more geometric way to see this is to
consider the blowdown of the various curves. For a $-n$ curve that abuts a
$-1$ curve, a blowdown of the $-1$ curve shifts the self-intersection as
$n\rightarrow(n-1)$. Thus, after blowdown we get:%
\begin{equation}
\mathcal{C}=13131\rightarrow1221\rightarrow 11 \rightarrow 0,
\end{equation}
a contradiction.

There is, however, a simple way to arrange a specific class of graphs with a
positive definite intersection form: These are given by taking an ADE\ graph
and simply making the self-intersection of a given curve more negative. We
will shortly put this observation to use.

\section{Classification of Minimal Models \label{sec:ENDPOINT}}

As should now be clear, there is a rich class of 6D\ SCFTs which can be
realized in F-theory. To classify the different possibilities, we need a
systematic way to generate SCFTs, and moreover, a way to specify when two
singular F-theory geometries yield the same conformal fixed point.

So, suppose we have an F-theory model with some choice of smooth base $B$.
After all resolutions in the base have been performed, we have reached the
Coulomb branch of some 6D\ SCFT. We would like to know which SCFT\ this is
associated with after shrinking all curves in the base to zero size.

As a first step in our classification program, we can consider the related
singular geometry obtained by blowing down all $-1$ curves of the
configuration of curves in $B$. This corresponds to eliminating the blown up
$-1$ curve from the configuration. When we do this, the self-intersection of
the remaining curves intersecting each such $-1$ curve will also shift. For
example, if we have a curve $\Sigma$ of self-intersection $-n$ which
intersects the $-1$ curve $E$, the class of the curve $\Sigma$ will shift as:
\begin{equation}
\lbrack\Sigma]\rightarrow\lbrack\Sigma]+ [E]=[\Sigma_{\text{new}}]
\end{equation}
and the self-intersection of $[\Sigma_{\text{new}}]$ is now $-(n-1)$ since:%
\begin{equation}
\lbrack\Sigma_{\text{new}}]\cdot\lbrack\Sigma_{\text{new}}]=[\Sigma
]\cdot\lbrack\Sigma]+2[\Sigma]\cdot [E]+[E]\cdot [E]=-(n-1).
\end{equation}
The original configuration of curves $\mathcal{C}$ will therefore blow down to
a new configuration $\mathcal{C}^{\prime}$. In this new configuration of
curves, the process of blowdown may create new $-1$ curves. Blowing down all
of the new $-1$ curves, and iterating this process, we eventually reach a
configuration of curves which contains no $-1$ curves. We shall refer to this
as the \textquotedblleft endpoint\textquotedblright\ of a configuration of
curves, $\mathcal{C}\rightarrow...\rightarrow\mathcal{C}_{\text{end}}$, and
the corresponding base as $B_{\text{end}}$. Examples of such endpoints include
the base $B=\mathbb{C}^{2}$, and the non-Higgsable clusters.

The main result of this section and section \ref{sec:LIST} will be a classification of \textit{all} such
endpoints. We shall refer to the corresponding six-dimensional SCFTs as ``minimal'' because all of the
other F-theory CFTs are obtained by supplementing these theories by additional ingredients such as further blowups and/or by
making the elliptic fibration more singular.

Quite surprisingly, it turns out that all the endpoints are specified by taking F-theory on a base of the form:
\begin{equation}
B_{\text{end}}=\mathbb{C}^{2}/\Gamma,
\end{equation}
where $\Gamma$ is a discrete subgroup of $U(2)$. For each consistent endpoint, there is a minimal 6D SCFT.
The full list of consistent F-theory model bases includes $\Gamma$ an ADE\ subgroup of $SU(2)$, as well as
specific discrete subgroups of $U(2)$ which we label as%
\begin{align}
A(x_{1},...,x_{r})\text{ \ \ for \ \ }\mathcal{C}_{\text{end}}  &
=x_{1}...x_{r}\\
D(y|x_{1},...,x_{\ell})\text{ \ \ for \ \ }\mathcal{C}_{\text{end}}  & =%
\begin{tabular}
[c]{|l|l|l|}\hline
& $2$ & \\\hline
$2$ & $y$ & $x_{1}...x_{\ell}$\\\hline
\end{tabular}
\end{align}
The group $A(x_{1},...,x_{r})$ is cyclic of order $p$ with generator acting on
the $\mathbb{C}^{2}$ coordinates $(z_{1},z_{2})$ by:
\begin{equation}
(z_{1},z_{2})\mapsto(\omega z_{1},\omega^{q}z_{2})\text{, \ \ where
\ \ }\omega=e^{2\pi i/p}\text{\ \ and \ \ }\frac{p}{q}=x_{1}-\frac{1}%
{x_{2}-...\frac{1}{x_{r}}}.
\end{equation}
The resolution of singularities in this case is a chain $x_{1}x_{2}\cdots
x_{r}$. Note that reading from right to left, i.e. taking the mirror image
configuration $x_r ,..., x_1 $ leads to a different generator of the same orbifold group.

There is also a discrete subgroup of $U(2)$ associated with the configuration of curves $D(y|x_{1},...,x_{\ell})$.
In this case, the group is a generalized D-type group, and is often labelled as $D_{p+q,q}$ where $p$ and $q$ are associated with
the Hirzebruch-Jung continued fraction:
\begin{equation}\label{Dorbo}
\frac{p}{q}=(y-1) -\frac{1}%
{x_{1}-...\frac{1}{x_{\ell}}}.
\end{equation}
For the present work we will not need the explicit form of this group. For further discussion
see e.g., reference \cite{UTWOorb},\footnote{We thank D.S. Park
for bringing this reference to our attention.} as well as \cite{Riemen:dvq, Brieskorn}. The
finite subgroups of $U(2)$ have been classified
\cite{MR1119304,MR0169108,MR2114077} and our groups $A(x_{1},...,x_{r})$ and
$D(y|x_{1},...,x_{\ell})$ do not exhaust the list, but in addition to
the ADE subgroups of $SU(2)$, they are the only ones which occur as endpoints in F-theory.

To reach this result, we shall first show that to be a consistent F-theory
model, the topology of any endpoint configuration has the branch structure of
an ADE\ graph, but where the self-intersection of the curves associated with
the nodes of the graph can be different from $-2$. In particular, we show that
if an endpoint contains a curve with self-intersection $-n$ with $n\geq3$,
then the branch structure of the graph is of A- or D-type.

We proceed by assuming that an endpoint exists, and then determine when such
endpoints can arise from a blowdown of a configuration of non-Higgsable
clusters. If the putative elliptic fibration over the
base becomes too singular, we will need to perform
some blowups in the base.\footnote{Although we are not discussing
Weierstrass equations here, the need to perform blowups can be
recognized directly from the equations:  any point at which $f$ has
multiplicity at least $4$ and $g$ has multiplicity at least $6$ must
be blown up.} For a curve of self intersection $-x$ for $x\geq1$, this
will involve blowing up the various intersection points with neighboring
curves. Each such blowup shifts the self-intersection of the original curve,
so after $j$ blowups, the self-intersection will be $-x^{(j)}$ where:%
\begin{equation}
x^{(j)}=x+j.
\end{equation}
On the other hand, since we must be able to resolve the base geometry to some
consistent gluing of non-Higgsable clusters, there are upper bounds on how
many times a given curve is allowed to be blown up. For example we must have
$x^{(j)}\leq12$.

As we have already mentioned, the first important simplification is that any consistent graph is a tree-like structure,
i.e., it contains no closed loops. This is explained in detail in
Appendix~\ref{APP:constraints}. Restricting our analysis to tree-like structures, the remainder of our
analysis proceeds in stages. First, we state the general procedure for how to
check whether an endpoint can serve as a base for an F-theory model. Then, we whittle
away at the possibilities, first by showing that no quartic vertices can exist
in an endpoint configuration. Then, we show that there can be at most one
trivalent vertex, and that such a vertex can only appear at the end of a
graph. This will establish the main claim that all consistent endpoints are
given by $U(2)$ orbifolds of $\mathbb{C}^{2}$ which form generalizations of
the A- and D-type series of Dynkin diagrams. We follow this in section \ref{sec:LIST}
with a list of all consistent endpoints.

\subsection{Algorithm for Minimal Resolutions \label{ssec:Algorithm}}

An important feature of each endpoint is that the intersection pattern
dictates a unique minimal resolution of the base geometry. For the ADE\ series
of the $(2,0)$ theories, no further resolution is required because these are
already non-Higgsable clusters. However, the seemingly innocuous operation of
modifying even a single self-intersection in such a graph can trigger a large
set of additional resolutions to reach a base which supports a
nonsingular F-theory model.  In other words, \textquotedblleft hidden
between the cracks\textquotedblright\ of an endpoint configuration
are a large set of additional curves.

The main point is that for a given endpoint $\mathcal{C}_{\text{end}}$, there
is an algorithm for reaching the uniquely specified minimal resolution:

\begin{itemize}
\item I) Check all pairs of nearest neighbors, $x_{i}x_{i+1}$. If any pairwise
intersection occurs which is not on our list of possibilities for a
non-Higgsable cluster, we need to blow up the intersection point, i.e., go to
$x_{i}^{(1)}1x_{i+1}^{(1)}$. Iterate throughout the entire graph.

\item II)\ Next, check that the gauging condition is satisfied, i.e., that for
any pair of curves separated by a $-1$ curve, that the associated gauge
algebra $\mathfrak{g}\oplus\mathfrak{g}^{\prime}\subset e_{8}$.
(The fact that this is the correct condition when the F-theory model
is maximally Higgsed is verified geometrically
in Appendix~\ref{APP:gluing}.)  If it is
satisfied, stop. Otherwise, blowup further.

\item III)\ In such cases, one must always blowup the intersection of the $-1$
and $-n$ curve for $n>3$, and for $2\leq n \leq3$, it will depend on whether
these are part of a non-Higgsable cluster. If they are part of a non-Higgsable
cluster, then no blowup is needed. If they are not part of a non-Higgsable
cluster, then blowup this intersection point as well.

\item IV)\ Keep repeating until one reaches a configuration of non-Higgsable
clusters connected by $-1$ curves.
\end{itemize}

In our \texttt{arXiv} submission we also include a \texttt{Mathematica} notebook which
automates this procedure. See Appendix \ref{APP:mathematica} for details.

\begin{figure}[ptb]
\begin{center}
\includegraphics[
height=1.5662in,
width=4.2722in
]{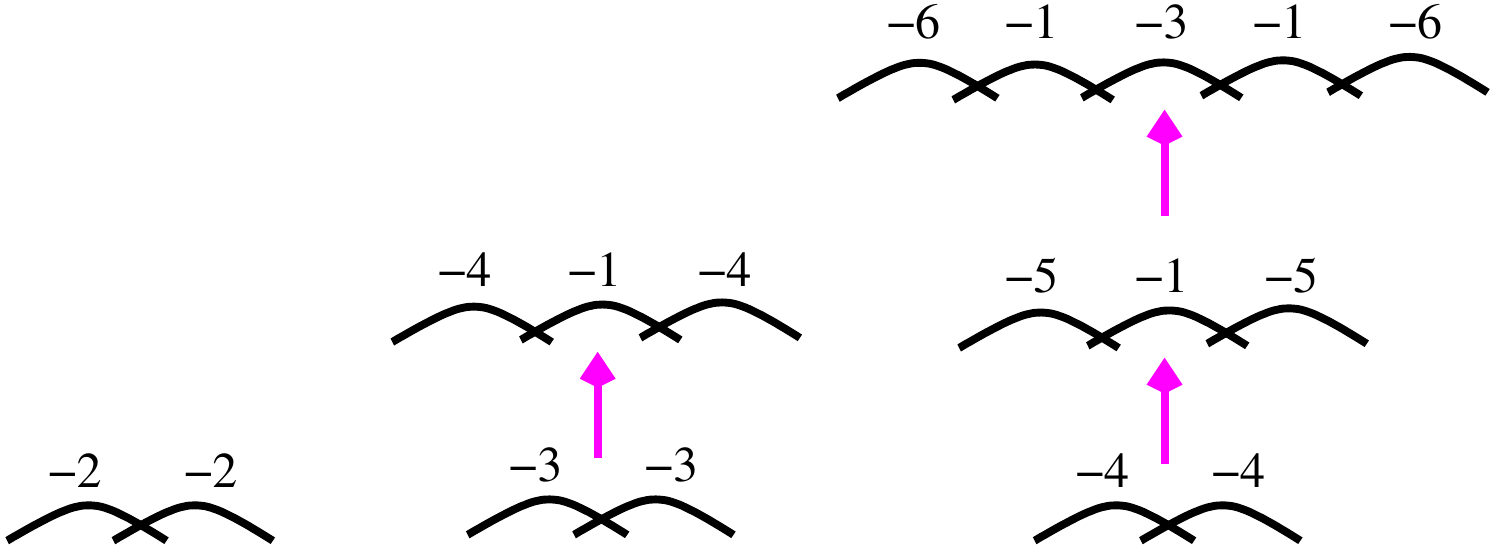}
\end{center}
\caption{Depiction of the minimal resolution associated with a given endpoint
configuration. For a pair of $-2$ curves which intersect, we have the $(2,0)$
theory $A_{2}$. For a pair of $-3$ curves which intersect, a single resolution
is required, after which we reach a consistent configuration of non-Higgsable
clusters. For a pair of $-4$ curves, additional resolutions are required. For
further discussion of these examples, see lines (\ref{33first})-(\ref{45last}%
).}%
\label{examplenew}%
\end{figure}

It is helpful to illustrate how to perform such resolutions with some examples
(see figure \ref{examplenew} for a depiction). Consider the $(1,0)$ theory
with endpoint:%
\begin{equation}
\mathcal{C}_{\text{end}}=33. \label{33first}%
\end{equation}
By inspection, this is not a non-Higgsable cluster, so we need to blowup the
intersection point between the two $-3$ curves. Doing this blowup shifts the
self-intersection of the two curves, so we reach:%
\begin{equation}
\mathcal{C}_{\text{end}}\rightarrow414.
\end{equation}
Now, we see that each $-4$ curve supports an $\mathfrak{so}(8)$ gauge
symmetry. We now check the gauging condition at the pair of curves touching
our $-1$ curve. Since $\mathfrak{so}(8) \oplus \mathfrak{so}(8)\subset
\mathfrak{e}_{8}$, we can stop blowing up.

As another example, consider the endpoint configuration:%
\begin{equation}
\mathcal{C}_{\text{end}}=44.
\end{equation}
Following the rules of our algorithm, we first note that this configuration is
not a non-Higgsable cluster, so we need to blowup the pair to reach:%
\begin{equation}
\mathcal{C}_{\text{end}}\rightarrow515.
\end{equation}
Now, we check the gauging condition. Here, each $-5$ curve supports an $\mathfrak{f}_{4}$
gauge algebra, so because the product $\mathfrak{f}_{4} \oplus \mathfrak{f}_{4}$
is not contained in $\mathfrak{e}_{8}$, a further blowup is required. In fact,
each intersection of the $-1$ curve with the $-5$ curve needs to be blown up.
First blowup the leftmost $-5$ curve:%
\begin{equation}
\mathcal{C}_{\text{end}}\rightarrow6125,
\end{equation}
and we see that the sequence $25$ is also not a non-Higgsable cluster. So, we
need to blowup the intersection between the $-2$ and $-5$ curve to reach:%
\begin{equation}
\mathcal{C}_{\text{end}}\rightarrow61316.
\end{equation}
Now we have a set of non-Higgsable clusters, so we again need to check the
gauging condition. Here, we have a $-6$ curve which supports an $\mathfrak{e}%
_{6}$ gauge algebra, and the $-3$ curve supports an $\mathfrak{su}(3)$ gauge
algebra, so we can stop blowing up.

As a final example, consider the endpoint configuration:%
\begin{equation}
\mathcal{C}_{\text{end}}=45.
\end{equation}
Again, the steps are rather similar. We blowup once to reach:%
\begin{equation}
\mathcal{C}_{\text{end}}\rightarrow516.
\end{equation}
Inspection of the clusters yields the minimal algebra $\mathfrak{f}_{4} \oplus \mathfrak{e}_{6}$,
which is not a subalgebra of $\mathfrak{e}_{8}$. So, we need to blowup at the intersection of
$-5$ and $-1$, as well as at the intersection of $-1$ and $-6$ to reach:
\begin{equation}
\mathcal{C}_{\text{end}}\rightarrow6126\rightarrow61317.
\end{equation}
We cannot stop here, however, because the $-7$ curve supports an
$\mathfrak{e}_{7}$ algebra, and the $-3$ curve supports an $\mathfrak{su}(3)$
algebra, so the product is not a subalgebra of $\mathfrak{e}_{8}$. Following
our algorithm, we first attempt blowing up at the intersection of the $-1$ and
$-7$ curve, so we reach:%
\begin{equation}
\mathcal{C}_{\text{end}}\rightarrow613218.
\end{equation}
Now, the $-8$ curve supports an $\mathfrak{e}_{7}$ algebra, and the $-2$ curve
supports an $\mathfrak{su}(2)$ algebra, so since
$\mathfrak{su}(2) \oplus \mathfrak{e}_{7}\subset\mathfrak{e}_{8}$, this satisfies the gauging
condition. But we are still not done, because now the pair given by the $-6$
curve and $-3$ curve do not satisfy the gauging condition: The $-6$ curve
supports an $\mathfrak{e}_{6}$ algebra, while the $-3$ curve now supports (via
the classification of algebras for non-Higgsable clusters) a $\mathfrak{g}%
_{2}$ algebra. So, we must at least blowup the intersection between the $-1$
curve and the $-6$ curve. Doing so, we get:
\begin{equation}
\mathcal{C}_{\text{end}}\rightarrow7123218, \label{45last}%
\end{equation}
and as we can verify, this is a configuration of non-Higgsable clusters where
the pairwise gauging condition is satisfied. All of the examples on our
endpoint list follow the same pattern, and can be worked out by applying the
same algorithm.

Let us give some additional more involved examples along these lines. Consider
the endpoint configuration of curves:%
\begin{equation}
\mathcal{C}_{\text{end}}=3A_{N}=3\underset{N}{\underbrace{2...2}}.
\end{equation}
For $N\geq3$, this is not a non-Higgsable cluster, and requires a blowup at
the intersection of the $-3$ and $-2$ curve. Doing so yields a new
configuration:%
\begin{equation}
\mathcal{C}_{\text{end}}\rightarrow413\underset{N-1}{\underbrace{2...2}}.
\end{equation}
If $N-1=2$, we can stop blowing up, but if $N-1>2$, then a further set of
blowups is required. As a consequence, there is a further cascade of blowups,
and we eventually reach the minimal resolution:%
\begin{equation}
\mathcal{C}_{\text{end}}\rightarrow\underset{N-2}{\underbrace{41...41}}322,
\end{equation}
where in the above, the pattern \textquotedblleft$41$\textquotedblright%
\ repeats $N-2$ times.

Another general feature of our classification is that all of the infinite
families mostly consist of $-2$ curves. The reason for this is that a cascade
of blowups rapidly shifts the value of the self-intersection closer to $-12$, and this is a
strict lower bound which cannot be violated. To illustrate, consider the endpoint
configuration:%
\begin{equation}
\mathcal{C}_{\text{end}}=7A_{N}7=7\underset{N}{\underbrace{2...2}}7,
\end{equation}
for $N$ sufficiently large. After performing all blowups necessary to reach a
smooth F-theory model, we reach a configuration with many additional curves:%
\begin{equation}
\mathcal{C}_{\text{end}}\rightarrow(12)12231513221\underset{N}{\underbrace
{(12)122315...513221(12)}}12231513221(12), \label{bigboy}%
\end{equation}
where in the end configuration there are $N+2$ curves with self-intersection
$-12$, i.e., the $N$ original $-2$ curves as well as the two $-7$ curves.

\subsection{No Quartic Vertices}

We now show that in an endpoint configuration of curves, any given curve can
only intersect at most three other curves. To show this, suppose to the
contrary, i.e., there is a curve in an endpoint configuration which intersects
at least four other curves. First of all, we see that such a graph must
contain at least one curve with self-intersection $-n$ for $n\geq3$.
Otherwise, we would have a graph made with just $-2$ curves, and there is no
ADE\ Dynkin diagram with a quartic vertex.

Now, to reach a non-singular F-theory model, we will need to blowup the
various intersection points. Without loss of generality, we can focus on a
subgraph such as:%
\begin{equation}
\mathcal{C}_{\text{sub}}=%
\begin{tabular}
[c]{|l|l|l|}\hline
& $y_{1}$ & \\\hline
$y_{4}$ & $x$ & $y_{2}$\\\hline
& $y_{3}$ & \\\hline
\end{tabular}
\ \ \ \ \ ,
\end{equation}
with at least one curve not a $-2$ curve. Now, this requires us to blowup at
least two intersections of $x$ with its neighbors. In fact, once we perform
two blowups on $x$, the self-intersection of this curve is high enough that the
other intersection points must also be blown up. By a similar token, this
forces even more blowups:
\begin{equation}
\mathcal{C}_{\text{sub}}\ \ \ \rightarrow%
\begin{tabular}
[c]{|l|l|l|l|l|}\hline
&  & $y_{1}^{(1)}$ &  & \\\hline
&  & $1$ &  & \\\hline
$y_{4}^{(1)}$ & $1$ & $x^{(4)}$ & $1$ & $y_{2}^{(1)}$\\\hline
&  & $1$ &  & \\\hline
&  & $y_{3}^{(1)}$ &  & \\\hline
\end{tabular}
\rightarrow%
\begin{tabular}
[c]{|l|l|l|l|l|l|l|}\hline
&  &  & $y_{1}^{(1)}$ &  &  & \\\hline
&  &  & $2$ &  &  & \\\hline
&  &  & $1$ &  &  & \\\hline
$y_{4}^{(1)}$ & $2$ & $1$ & $x^{(8)}$ & $1$ & $2$ & $y_{2}^{(1)}$\\\hline
&  &  & $1$ &  &  & \\\hline
&  &  & $2$ &  &  & \\\hline
&  &  & $y_{3}^{(1)}$ &  &  & \\\hline
\end{tabular}
.
\end{equation}

But now the gauge algebra for $x^{(8)}$ is $\mathfrak{e}_{8}$, so we need four
more blowups, so we reach a curve $x^{(12)}>12$, a contradiction. We therefore
conclude that no quartic vertices can appear.

\subsection{Restrictions on Trivalent Vertices}

In fact, we can also show that in all graphs, there can be at most one
trivalent vertex, and aside from the E-type Dynkin diagrams, a trivalent
vertex can only appear in a generalized D-type graph.

The analysis proceeds along the same lines as for the exclusion of quartic
vertices: We assume the existence of a subgraph, and then perform blowups to
attempt to reach a consistent F-theory base consisting of non-Higgsable
clusters. If this is not possible, we reach a contradiction.

So to begin, suppose we have a subgraph containing a configuration of curves:%
\begin{equation}
\mathcal{C}_{\text{sub}}=%
\begin{tabular}
[c]{|l|l|l|}\hline
& $y_{1}$ & \\\hline
$y_{3}$ & $x$ & $y_{2}$\\\hline
\end{tabular}
\ \ \ \ \ \ .\label{trivalent}%
\end{equation}
We claim that at least two of the $y$'s must have self-intersection $-2$. That
is, the trivalent vertices are of the form:%
\begin{equation}
\mathcal{C}_{\text{sub}}=%
\begin{tabular}
[c]{|l|l|l|}\hline
& $2$ & \\\hline
$2$ & $x$ & $y$\\\hline
\end{tabular}
\ \ \ \ \ \ .
\end{equation}
To see how this restriction comes about, suppose to the contrary that at least
two of the $y_{i}$'s have self-intersection $-3$ or less. Without loss of
generality, take them to be $y_{1}$ and $y_{3}$. It becomes necessary to
blowup all intersections of the $x$ curve with the $y$ curves. Doing so, one
reaches the sequence of configurations:%
\begin{equation}
\mathcal{C}_{\text{sub}}\rightarrow%
\begin{tabular}
[c]{|l|l|l|l|l|}\hline
&  & $y_{1}^{(1)}$ &  & \\\hline
&  & $1$ &  & \\\hline
$y_{3}^{(1)}$ & $1$ & $x^{(3)}$ & $1$ & $y_{2}^{(1)}$\\\hline
\end{tabular}
\rightarrow%
\begin{tabular}
[c]{|l|l|l|l|l|l|l|l|}\hline
&  &  &  & $y_{1}^{(2)}$ &  &  & \\\hline
&  &  &  & $1$ &  &  & \\\hline
&  &  &  & $3$ &  &  & \\\hline
&  &  &  & $1$ &  &  & \\\hline
$y_{3}^{(2)}$ & $1$ & $3$ & $1$ & $x^{(6)}$ & $1$ & $2$ & $y_{2}^{(1)}%
$\\\hline
\end{tabular}
\ \ \ \ \ .
\end{equation}
Now, at this stage, $x^{(6)}\geq8$, so the gauge algebra of the $x$-curve is
at least $\mathfrak{e}_{7}$, and the gauging condition with each of the $-3$
curves is violated. Observe, however, that after even two more blowups, we
reach an $\mathfrak{e}_{8}$ gauge algebra on the $x$ curve. That means at
least five more blowups must be performed, so we reach the configuration:%
\begin{equation}
\mathcal{C}_{\text{sub}}\rightarrow%
\begin{tabular}
[c]{|l|l|l|l|l|l|l|l|l|l|l|}\hline
&  &  &  &  &  & $y_{1}^{(2)}$ &  &  &  & \\\hline
&  &  &  &  &  & $1$ &  &  &  & \\\hline
&  &  &  &  &  & $3$ &  &  &  & \\\hline
&  &  &  &  &  & $2$ &  &  &  & \\\hline
&  &  &  &  &  & $2$ &  &  &  & \\\hline
&  &  &  &  &  & $1$ &  &  &  & \\\hline
$y_{3}^{(2)}$ & $1$ & $3$ & $2$ & $2$ & $1$ & $x^{(11)}$ & $1$ & $2$ & $2$ &
$y_{2}^{(1)}$\\\hline
\end{tabular}
\ \ \ \ \ .
\end{equation}
This violates the condition $x^{(11)}\leq12$, a contradiction. In fact, the
same sort of argument reveals that the only trivalent vertices which can occur
are:%
\begin{equation}
\mathcal{C}_{\text{sub}}=%
\begin{tabular}
[c]{|l|l|l|}\hline
& $2$ & \\\hline
$2$ & $3$ & $2$\\\hline
\end{tabular}
\text{ \ \ or \ \ }%
\begin{tabular}
[c]{|l|l|l|}\hline
& $2$ & \\\hline
$2$ & $2$ & $y$\\\hline
\end{tabular}
\ \ \ \ \ ,
\end{equation}
where in the former case, this is actually the full endpoint.

We also learn that a trivalent vertex can only appear at the end of a graph,
i.e., at most one leg extends to a longer chain. Indeed, suppose to the
contrary that $y_{2}$ and $y_{3}$ of the graph in line (\ref{trivalent})
intersect some other curves with self-intersections $-z_{2}$ and $-z_{3}$:
\begin{equation}
\mathcal{C}_{\text{sub}}=%
\begin{tabular}
[c]{|l|l|l|l|l|}\hline
&  & $y_{1}$ &  & \\\hline
$z_{3}$ & $y_{3}$ & $x$ & $y_{2}$ & $z_{2}$\\\hline
\end{tabular}
\ \ \ \ \ \ .
\end{equation}
Since we are assuming that this is not a subgraph in a Dynkin diagram, there
will be at least one forced blowup, which will eventually trickle down to our
subgraph. Since each additional blowup adds additional constraints, it is
enough to demonstrate a contradiction if any of the curves in our subgraph has
self intersection $-n$ for $n>2$.

To this end, we can simply list all possible tree-like subgraphs with low
values of the $y$'s and $z$'s and check that this never leads to a smooth
F-theory model. For example, we have the sequence of forced blowups:%
\begin{equation}
\mathcal{C}_{\text{sub}}=%
\begin{tabular}
[c]{|l|l|l|l|l|}\hline
&  & $2$ &  & \\\hline
$3$ & $2$ & $2$ & $2$ & $2$\\\hline
\end{tabular}
\rightarrow...\rightarrow%
\begin{tabular}
[c]{|l|l|l|l|l|l|l|l|l|l|l|l|l|l|}\hline
&  &  &  &  &  &  &  & $3$ &  &  &  &  & \\\hline
&  &  &  &  &  &  &  & $2$ &  &  &  &  & \\\hline
&  &  &  &  &  &  &  & $2$ &  &  &  &  & \\\hline
&  &  &  &  &  &  &  & $1$ &  &  &  &  & \\\hline
$4$ & $1$ & $5$ & $1$ & $3$ & $2$ & $2$ & $1$ & $(12)$ & $1$ & $2$ & $2$ & $3$
& $2$\\\hline
\end{tabular}
\ \ \ \ ,
\end{equation}
which eventually leads to a contradiction. The other case to cover is:%
\begin{equation}
\mathcal{C}_{\text{sub}}=%
\begin{tabular}
[c]{|l|l|l|l|l|}\hline
&  & $3$ &  & \\\hline
$2$ & $2$ & $2$ & $2$ & $2$\\\hline
\end{tabular}
\rightarrow...\rightarrow%
\begin{tabular}
[c]{|l|l|l|l|l|l|l|l|l|l|l|}\hline
&  &  &  &  & $5$ &  &  &  &  & \\\hline
&  &  &  &  & $1$ &  &  &  &  & \\\hline
&  &  &  &  & $3$ &  &  &  &  & \\\hline
&  &  &  &  & $2$ &  &  &  &  & \\\hline
&  &  &  &  & $2$ &  &  &  &  & \\\hline
&  &  &  &  & $1$ &  &  &  &  & \\\hline
$2$ & $3$ & $2$ & $2$ & $1$ & $(12)$ & $1$ & $2$ & $2$ & $3$ & $2$\\\hline
\end{tabular}
\ \ \ \ ,
\end{equation}
which also leads to an inconsistent model. Putting a $-3$ curve further into
the interior only makes the putative elliptic fibration more singular.

As a result of these limitations on trivalent vertices, we can also see that
the short legs of the trivalent vertices must be $-2$ curves. Otherwise, we
will inflict too many blowups on the subgraph. For example, we reach a
contradiction for the configuration:%
\begin{equation}
\mathcal{C}_{\text{sub}}=%
\begin{tabular}
[c]{|l|l|l|l|}\hline
& $3$ &  & \\\hline
$2$ & $2$ & $2$ & $2$\\\hline
\end{tabular}
\rightarrow...\rightarrow%
\begin{tabular}
[c]{|l|l|l|l|l|l|l|l|l|l|}\hline
&  &  &  & $5$ &  &  &  &  & \\\hline
&  &  &  & $1$ &  &  &  &  & \\\hline
&  &  &  & $3$ &  &  &  &  & \\\hline
&  &  &  & $2$ &  &  &  &  & \\\hline
&  &  &  & $2$ &  &  &  &  & \\\hline
&  &  &  & $1$ &  &  &  &  & \\\hline
$3$ & $2$ & $2$ & $1$ & $(12)$ & $1$ & $2$ & $2$ & $3$ & $2$\\\hline
\end{tabular}
\ \ \ ,
\end{equation}
which also leads to an inconsistent model.

Hence, if a trivalent vertex appears in an endpoint, then it must appear as
part of a graph:%
\begin{equation}
\mathcal{C}_{\text{sub}}=%
\begin{tabular}
[c]{|l|l|l|l|}\hline
& $2$ &  & \\\hline
$2$ & $2$ & $y_{1}$ & $...$\\\hline
\end{tabular}
\ \ \ \ \ \ ,
\end{equation}
where nothing else intersects the $-2$ curves on the legs of the vertex. Due
to this limitation, we have already placed a strong restriction on endpoints:
There can be at most two trivalent vertices, and these can only appear at the
ends of a linear chain of curves.

In fact, there can be at most one trivalent vertex in an endpoint
configuration. Suppose to the contrary that we have a configuration of curves
with two trivalent vertices:%
\begin{equation}
\mathcal{C}_{\text{end}}=%
\begin{tabular}
[c]{|l|l|l|l|l|l|l|}\hline
& $2$ &  &  &  & $2$ & \\\hline
$2$ & $2$ & $y_{1}$ & $...$ & $y_{r}$ & $2$ & $2$\\\hline
\end{tabular}
\ \ \ \ \ ,
\end{equation}
where $r\geq1$. We must have at least one curve with self-intersection less
than $-2$, since the Dynkin diagram classification contains no graphs with two
trivalent vertices. By inspection, each pairwise intersection of curves must
be blown up at least once. This means we reach the configuration:%
\begin{equation}
\mathcal{C}_{\text{end}}\rightarrow%
\begin{tabular}
[c]{|l|l|l|l|l|l|l|l|l|l|l|}\hline
& $2$ &  &  &  &  &  &  &  & $2$ & \\\hline
$2$ & $3$ & $1$ & $y_{1}^{(2)}$ & $1$ & $...$ & $1$ & $y_{r}^{(2)}$ & $1$ &
$3$ & $2$\\\hline
\end{tabular}
\ \ \ \ \ .
\end{equation}
Because the $-3$ curves in each trivalent vertex sit in the middle of an NHC,
each supports an $\mathfrak{so}(7)$ gauge algebra. Since $\mathfrak{so}%
(7)\oplus\mathfrak{f}_{4}$ is not a subalgebra of $\mathfrak{e}_{8}$ we see
that if $y_{1}^{(2)}\geq3$, a further set of blowups will be generated. For
example, if $y_{1}^{(2)}\geq3$, then, there are some forced blowups:%
\begin{align}
\mathcal{C}_{\text{end}} &  \rightarrow%
\begin{tabular}
[c]{|l|l|l|l|l|l|l|l|l|l|l|}\hline
& $2$ &  &  &  &  &  &  &  & $2$ & \\\hline
$2$ & $3$ & $1$ & $y_{1}^{(2)}$ & $1$ & $...$ & $1$ & $y_{r}^{(2)}$ & $1$ &
$3$ & $2$\\\hline
\end{tabular}
\\
&  \rightarrow%
\begin{tabular}
[c]{|l|l|l|l|l|l|l|l|l|l|l|l|l|}\hline
& $2$ &  &  &  &  &  &  &  &  &  & $2$ & \\\hline
$2$ & $4$ & $1$ & $3$ & $1$ & $y_{1}^{(3)}$ & $1$ & $...$ & $1$ & $y_{r}%
^{(2)}$ & $1$ & $3$ & $2$\\\hline
\end{tabular}
\\
&  \rightarrow%
\begin{tabular}
[c]{|l|l|l|l|l|l|l|l|l|l|l|l|l|l|}\hline
&  & $3$ &  &  &  &  &  &  &  &  &  &  & \\\hline
&  & $1$ &  &  &  &  &  &  &  &  &  & $2$ & \\\hline
$3$ & $1$ & $6$ & $1$ & $3$ & $1$ & $y_{1}^{(3)}$ & $1$ & $...$ & $1$ &
$y_{r}^{(2)}$ & $1$ & $3$ & $2$\\\hline
\end{tabular}
\ \ \ .
\end{align}
Owing to the trivalent vertex on the righthand side, we now reach a
contradiction. The reason is that at least one more blowup on $y_{1}$ is
required, and this sets off a cascade on the leftmost vertex.

Based on this, we conclude that $y_{1}=y_{r}=2$. In fact, the same reasoning
applies for all of the interior curves, so we require $y_{i}=2$ for all $i$.
In other words, the only option left is an affine D-type Dynkin diagram. This,
however, contradicts the original requirement that we can simultaneously
contract all curves to zero size. We therefore conclude that at most one
trivalent vertex can appear in a graph.

\section{Endpoint Classification} \label{sec:LIST}

In this section we present the results of our classification of endpoints.
There are two steps to this procedure. First, we identify candidate endpoint
configurations consisting of generalized A- or D-type graphs. Then, we
perform a minimal resolution according to the algorithm detailed in
subsection \ref{ssec:Algorithm} (see also Appendix \ref{APP:mathematica}). Provided
we reach a set of non-Higgsable clusters, we admit this endpoint. Otherwise,
we reject it. The plan is simply to sweep over all possible endpoint
configurations using an automated program.

In fact, it is \textquotedblleft only\textquotedblright\ necessary to scan
configurations with ten curves or less. The reason is that starting at
eleven curves, the innermost curves of an endpoint are always $-2$ curves. This means that
any candidate patterns found at ten curves can either continue to a string of eleven or more curves, or go extinct. These patterns
are given by five curves on the left of a graph, and five on the right, with a string of $-2$ curves in the middle.

To see how this comes about, suppose we have a subgraph of the form:%
\begin{equation}
\mathcal{C}_{\text{sub}}=x_{1}x_{2}x_{3}x_{4}x_{5}x_{6}x_{7}x_{8}x_{9}%
x_{10}x_{11}.
\end{equation}
In this case, the claim is that $x_{6}=2$. Indeed, suppose to the contrary.
Then, we have the sequence of forced blowups:%
\begin{align}
\mathcal{C}_{\text{sub}}  &  \rightarrow x_{1}x_{2}x_{3}^{(1)}1x_{4}^{(2)}1x_{5}%
^{(2)}1x_{6}^{(2)}1x_{7}^{(2)}1x_{8}^{(2)}1x_{9}^{(1)}x_{10}x_{11}\\
&  \rightarrow x_{1}x_{2}x_{3}^{(1)}1x_{4}^{(3)}131x_{5}^{(4)}131x_{6}%
^{(4)}131x_{7}^{(4)}131x_{8}^{(3)}1x_{9}^{(1)}x_{10}x_{11}\\
&  \rightarrow x_{1}x_{2}x_{3}^{(1)}1x_{4}^{(3)}131x_{5}^{(4)}13221x_{6}%
^{(8)}12231x_{7}^{(4)}131x_{8}^{(3)}1x_{9}^{(1)}x_{10}x_{11}\\
&  \rightarrow x_{1}x_{2}x_{3}^{(1)}1x_{4}^{(3)}131x_{5}^{(5)}131513221x_{6}%
^{(10)}122315131x_{7}^{(5)}131x_{8}^{(3)}1x_{9}^{(1)}x_{10}x_{11},
\end{align}
so for $x_{6}>2$ we generate a contradiction. This means that starting at eleven or more curves,
the only configurations are:
\begin{equation}
\mathcal{C}_{\text{sub}}=x_{1}x_{2}x_{3}x_{4}x_{5}\underset{N}{\underbrace
{2...2}}y_{5}y_{4}y_{3}y_{2}y_{1}.
\end{equation}
Thus, classification requires us to determine all consistent $x_i$ and $y_j$.

There are some additional simplifications which help in reducing the time of
the computation. For two or more curves, the value of $x$ for a curve of self-intersection $-x$
is bounded as $2 \leq x \leq 9$. Moreover, the deeper we go into the interior
of a graph, the tighter the upper bound on $x$.

Rather than listing all of the different endpoints, we only give the ones
which are \textquotedblleft rigid\textquotedblright\ in the sense that any
further increase in the value of $x$ for a curve of self-intersection $-x$
would produce an inconsistent endpoint. Since decreasing the value of $x$
also produces a consistent endpoint\footnote{This is because a $-1$ curve
can meet the curve of self-intersection $-x$ in a larger configuration,
and blowing that $-1$ curve down yields a configuration in which $x$ has
been decreased by $1$.} (provided the new value is greater than or equal to two),
this collection of theories forms a convex hull, and our task is to determine
its shape.

For the A-type and D-type theories, we find the following infinite rigid
families:%
\begin{align}
\mathcal{C}_{\text{rigid}}^{(A)} &  =\alpha A_{N}\beta\text{ \ \ for
\ \ }N\geq0\\
\mathcal{C}_{\text{rigid}}^{(D)} &  =D_{N}\gamma\text{ \ \ \ \ for \ \ }N\geq2
\end{align}
where $A_{N}$ and $D_{N}$ refer to a configuration of $-2$ curves, which are
decorated on the end by configurations $\alpha$, $\beta$ and $\gamma$ of up to
five curves:%
\begin{align}
\alpha &  \in\left\{  7,33,24,223,2223,22223\right\}  \\
\beta &  \in\left\{  7,33,42,322,3222,32222\right\}  \\
\gamma &  \in\left\{  24,32\right\}  .
\end{align}
This is the complete list of rigid theories for ten or more curves. In addition
to these infinite families, there are some outlier rigid theories
for the A-type graphs which occur at nine or fewer curves:
\begin{equation}
\mathcal{C}_{\text{rigid outliers}}^{(A)}\in\left\{
\begin{array}
[c]{l}%
(12)\text{, }92\text{, }83\text{, }822\text{, }8222\text{, }82222\text{, }\\
352\text{, }343\text{, }262\text{, }2522\text{, }25222\text{, }252222\text{, }\\
22422\text{, }224222\text{, }2242222\text{, }2224222\text{, }22242222\text{, }222242222
\end{array}
\right\}.
\end{equation}
In the above list, the configuration $(12)$ refers to a single curve of
self-intersection $-12$. The mirror image configurations of curves
define a different generator of the same orbifold group.

In the rest of this section we give more details of performing such a sweep.
First, we give the generalized A-type graphs, and then the generalized D-type graphs. We then discuss
the characterization of some of these theories in terms of their associated orbifold group action.

\begin{table}[ptb]
\begin{center}
\begin{tabular}
[c]{|l|l|l|l|l|l|l|l|l|l|}\hline
$1$ & $2$ & $3$ & $4$ & $5$ & $6$ & $7$ & $8$ & $9$ & $10$\\\hline\hline
$(12)$ &  &  &  &  &  &  &  &  & \\\hline
& $92$ &  &  &  &  &  &  &  & \\\hline
& $83$ & $822$ & $8222$ & $82222$ &  &  &  &  & \\\hline
& $77$ & $742$ & $7322$ & $73222$ & $732222$ &  &  &  & \\\hline
&  & $733$ & $7233$ & $72322$ & $723222$ & $7232222$ &  &  & \\\hline
&  & $727$ & $7242$ & $72242$ & $722322$ & $7223222$ & $72232222$ &
$722232222$ & $7222232222$\\\hline
&  & $352$ & $7227$ & $72233$ & $722242$ & $7222322$ & $72223222$ &
$722223222$ & $7222223222$\\\hline
&  & $343$ & $3422$ & $72227$ & $722233$ & $7222242$ & $72222322$ &
$722222322$ & $7222222322$\\\hline
&  & $262$ & $3342$ & $34222$ & $722227$ & $7222233$ & $72222242$ &
$722222242$ & $7222222242$\\\hline
&  &  & $3333$ & $33322$ & $342222$ & $7222227$ & $72222233$ & $722222233$ &
$7222222233$\\\hline
&  &  & $2522$ & $33242$ & $333222$ & $3332222$ & $72222227$ & $722222227$ &
$7222222227$\\\hline
&  &  & $2442$ & $33233$ & $332322$ & $3323222$ & $33232222$ & $332232222$ &
$3322232222$\\\hline
&  &  &  & $2522$ & $332242$ & $3322322$ & $33223222$ & $332223222$ &
$3322223222$\\\hline
&  &  &  & $24322$ & $332233$ & $3322242$ & $33222322$ & $332222322$ &
$3322222322$\\\hline
&  &  &  & $24242$ & $252222$ & $3322233$ & $33222242$ & $332222242$ &
$3322222242$\\\hline
&  &  &  & $22422$ & $243222$ & $2432222$ & $33222233$ & $332222233$ &
$3322222233$\\\hline
&  &  &  &  & $242322$ & $2423222$ & $24232222$ & $242232222$ & $2422232222$%
\\\hline
&  &  &  &  & $242242$ & $2422322$ & $24223222$ & $242223222$ & $2422223222$%
\\\hline
&  &  &  &  & $224222$ & $2422242$ & $24222322$ & $242222322$ & $2422222322$%
\\\hline
&  &  &  &  & $223322$ & $2242222$ & $24222242$ & $242222242$ & $2422222242$%
\\\hline
&  &  &  &  &  & $2233222$ & $22332222$ & $223232222$ & $2232232222$\\\hline
&  &  &  &  &  & $2232322$ & $22323222$ & $223223222$ & $2232223222$\\\hline
&  &  &  &  &  & $2224222$ & $22322322$ & $223222322$ & $2232222322$\\\hline
&  &  &  &  &  &  & $22242222$ & $222332222$ & $2223232222$\\\hline
&  &  &  &  &  &  & $22233222$ & $222242222$ & $2222332222$\\\hline
\end{tabular}
\end{center}
\caption{Table of rigid theories for A-type graphs for up to ten curves. The first row indicates
the number of curves, and all subsequent rows indicate the specific endpoint configuration. In the first column,
$(12)$ refers to a single curve of self-intersection $-12$. In all other cases, the self-intersection of the curves is $-x$
for $2 \leq x \leq 9$. As the number of curves increases, some patterns either go extinct, or persist to an arbitrary
number of curves.}
\label{tab:ATYPE}%
\end{table}

\subsection{Generalized A-Type Theories}

We begin with A-type graphs of the form:%
\begin{equation}
\mathcal{C}_{\text{end}}=x_{1}...x_{r}.
\end{equation}
There is a built in symmetry here due to the fact that the word $x_{r}%
...x_{1}$ specifies the same geometry. We shall therefore
only give one representative in our endpoint list, sorting lexicographically
according to larger values of $x_{1}$ first. There are some clear patterns
which either go extinct, or continue on to an arbitrary number of curves. For the list of
rigid A-type theories with up to ten curves, see table \ref{tab:ATYPE}.

At this point, we can identify the general patterns which persist to an
arbitrary number of curves:
\begin{equation}
\mathcal{C}_{\text{rigid}}^{(A)}=\alpha A_{N}\beta\text{ \ \ for \ \ }N\geq0
\end{equation}
where $\alpha$ and $\beta$ are configurations of up to five curves in the set:%
\begin{align}
\alpha & \in\left\{  7,33,24,223,2223,22223\right\}  \\
\beta & \in\left\{  7,33,42,322,3222,32222\right\}  .
\end{align}
Additionally, table \ref{tab:ATYPE} contains all outlier cases for nine or fewer curves.

\subsection{Generalized D-Type Theories}

In the case of the D-type theories, we can perform a similar sweep. Due
to the presence of a trivalent vertex in D-type graphs, far fewer cases
need to be considered. For example, at four curves, we have the rigid
configurations:%
\begin{equation}
\mathcal{C}_{\text{rigid}}=%
\begin{tabular}
[c]{|l|l|l|}\hline
& $2$ & \\\hline
$2$ & $3$ & $2$\\\hline
\end{tabular}
\ \ \text{, \ \ }%
\begin{tabular}
[c]{|l|l|l|}\hline
& $2$ & \\\hline
$2$ & $2$ & $4$\\\hline
\end{tabular}
\ \ .
\end{equation}
The minimal resolution for these two cases is:%
\begin{equation}%
\begin{tabular}
[c]{|l|l|l|}\hline
& $2$ & \\\hline
$2$ & $3$ & $2$\\\hline
\end{tabular}
\rightarrow%
\begin{tabular}
[c]{|l|l|l|l|l|}\hline
&  & $3$ &  & \\\hline
&  & $1$ &  & \\\hline
$3$ & $1$ & $6$ & $1$ & $3$\\\hline
\end{tabular}
.
\end{equation}
and:%
\begin{equation}%
\begin{tabular}
[c]{|l|l|l|}\hline
& $2$ & \\\hline
$2$ & $2$ & $4$\\\hline
\end{tabular}
\ \rightarrow%
\begin{tabular}
[c]{|l|l|l|l|}\hline
& $2$ &  & \\\hline
$2$ & $3$ & $1$ & $5$\\\hline
\end{tabular}
\rightarrow%
\begin{tabular}
[c]{|l|l|l|l|l|l|l|}\hline
&  & $3$ &  &  &  & \\\hline
&  & $1$ &  &  &  & \\\hline
$3$ & $1$ & $6$ & $1$ & $3$ & $1$ & $6$\\\hline
\end{tabular}
.
\end{equation}
The reason for the second blowup is that the $-5$ curve abuts a $-3$
curve in the middle of a non-Higgsable cluster, so the algebra
is $\mathfrak{f}_{4} \oplus \mathfrak{so}(7)$, which is
not a subalgebra of $\mathfrak{e}_{8}$. Compared with the A-type graphs, this
is the reason such graphs produce far fewer rigid theories. In the minimal
resolutions of the A-type graphs, a $-3$ curve which abuts a $-5$ curve yields
a $\mathfrak{g}_{2}$ algebra, and $\mathfrak{f}_{4} \oplus \mathfrak{g}_{2}$  is
a subalgebra of $\mathfrak{e}_{8}$.

At five curves and above, all of the graphs have the general form:%
\begin{equation}
\mathcal{C}_{\text{end}}=%
\begin{tabular}
[c]{|l|l|l|}\hline
& $2$ & \\\hline
$2$ & $2$ & $x_{1}...x_{r}$\\\hline
\end{tabular}
\ \ .
\end{equation}
In fact, we can simply repurpose our classification results for A-type graphs
for this case as well. Owing to the trivalent vertex, we cannot tolerate many
blowups in its vicinity. For rigid D-type theories, we find no outliers,
and all are of one of two types:%
\begin{equation}
\mathcal{C}_{\text{rigid}}^{(D)}=D_{N}\gamma\text{ \ \ for }N\geq2\text{,}%
\end{equation}
where $\gamma$ is a configuration of curves in the set:%
\begin{equation}
\gamma\in\left\{  32, 24 \right\}.
\end{equation}

\subsection{Orbifold Examples: A-type theories}
Since we now have a list of the various endpoints, we can also determine the
explicit form of the orbifold group action. In this subsection, we focus
on some of the A-type theories.

Recall that for an A-type graph of curves $x_{1}...x_{r}$, we have
the orbifold group action $A(x_{1},...,x_{r})$:%
\begin{equation}
(z_{1},z_{2})\mapsto(\omega z_{1},\omega^{q}z_{2})\text{ \ \ where
\ \ }\omega= e^{2\pi i/p}\text{
\ \ and \ \ }\frac{p}{q}=x_{1}-\frac{1}{x_{2}-...-\frac{1}{x_{r}}}.
\end{equation}
In table \ref{tab:RIGIDorbs} we collect the values of $p/q$ for all
rigid configurations of ten or more curves.
\begin{table}
\begin{center}
\renewcommand{\arraystretch}{2}
\begin{tabular}
[c]{|l|l|l|l|l|l|l|}\hline
& $22227$ & $22233$ & $22242$ & $22322$ & $23222$ & $32222$\\\hline
$72222$ & $\frac{36N+336}{6N+55}$ & $\frac{30N+264}{5N+43}$ & $\frac
{30N+257}{5N+42}$ & $\frac{24N+190}{4N+31}$ & $\frac{30N+209}{5N+34}$ &
$\frac{36N+216}{6N+35}$\\\hline
$33222$ &  & $\frac{25N+205}{10N+81}$ & $\frac{25N+200}{10N+79}$ &
$\frac{20N+147}{8N+58}$ & $\frac{25N+160}{10N+63}$ & $\frac{30N+163}{12N+64}%
$\\\hline
$24222$ &  &  & $\frac{25N+195}{15N+116}$ & $\frac{20N+143}{12N+85}$ &
$\frac{25N+155}{15N+92}$ & $\frac{30N+157}{18N+93}$\\\hline
$22322$ &  &  &  & $\frac{16N+104}{12N+77}$ & $\frac{20N+111}{15N+82}$ &
$\frac{24N+110}{18N+81}$\\\hline
$22232$ &  &  &  &  & $\frac{25N+115}{20N+91}$ & $\frac{30N+109}{24N+86}%
$\\\hline
$22223$ &  &  &  &  &  & $\frac{36N+96}{30N+79}$\\\hline
\end{tabular}
\end{center}
\caption{Table of values for the fraction $p/q=x_{1}-\frac{1}{x_{2}%
-...-\frac{1}{x_{r}}}$ for the $21$ physically distinct rigid endpoint
configurations at ten or more curves. All of these configurations are of the
form $\alpha A_{N}\beta$ for appropriate $\alpha$ and $\beta$ a configuration
of five curves indicated in the highest row and leftmost column. The
orbifold group action is $(z_{1},z_{2})\mapsto(\omega z_{1},\omega^{q}z_{2})$ for $\omega$ a primitive
$p^{\text{th}}$ root of unity. By reversing the order of the curves in the
graph, we would get a different generator of the orbifold group action which
is geometrically identical.}
\label{tab:RIGIDorbs}%
\end{table}

Consider next the $xA_{N}y$ theories for $2 \leq x , y \leq 7$. The orbifold generated by the $1/p$ and
$q/p$ roots of unity is:%
\begin{equation}
\frac{1}{p}=\frac{1}{N(x-1)(y-1)+xy-1}\text{ \ \ and \ \ }\frac{q}{p}%
=\frac{N(y-1)+y}{N(x-1)(y-1)+xy-1}.\label{qpdoll}%
\end{equation}
An interesting feature of this example is that at least for some of the rigid theories
with $x=y=7$, there is a simple realization of the F-theory model as an
orbifold $(\mathbb{C}^{2}\times T^{2})/\Gamma$:%
\begin{equation}
\Gamma:\left(  z_{1},z_{2},\lambda\right)  \mapsto\left(  \eta\zeta^{-1} z_{1}%
,\eta\zeta z_{2},\eta^{-2}\lambda\right)  ,
\end{equation}
where the group action in the base $\left(  z_{1},z_{2}\right)  \mapsto\left(
\eta\zeta^{-1} z_{1},\eta\zeta z_{2}\right)  $ is by an A-type orbifold. For
example, consider%
\begin{equation}
\eta=\exp\left(  2\pi i\cdot\frac{1}{12}\right)  \text{ \ \ and \ \ }%
\zeta=\exp\left(  2\pi i\cdot\frac{k}{12k+1}\right)  \text{.}%
\end{equation}
The base of the F-theory model is an A-type $U(2)$ orbifold of $\mathbb{C}%
^{2}$ with values of $1/p$ and $q/p$:%
\begin{equation}
\frac{1}{p}=\frac{1}{12(12k+1)}\text{ \ \ and \ \ }\frac{q}{p}=\frac
{24k+1}{12(12k+1)},
\end{equation}
which matches to our general expression in line (\ref{qpdoll}) for $N = 4 k - 1$
and $x=y=7$.

Finally, as a brief aside, let us note that one can also generate F-theory
models on $(\mathbb{C}^{2}\times T^{2})/\Gamma$ by combining an ADE\ orbifold
of $\mathbb{C}^{2}$ with a $\mathbb{Z}_{n}$ orbifold of the $T^{2}$:
\begin{equation}
\Gamma:\left(  z_{1},z_{2},\lambda\right)  \mapsto\left(  \alpha\rho_{1}%
(z_{1},z_{2}),\alpha\rho_{2}\left(z_{1} , z_{2}\right)  ,\alpha^{-2}\lambda\right)  ,
\end{equation}
where $\alpha^2$ is of order $1,2,3,4,6$. Here, $\rho_{i}$ denotes the group action on the $z_{i}$
coordinate and $\lambda$ is the holomorphic differential of a constant $T^{2}$.
The special case where the ADE subgroup is trivial corresponds to the class
of examples reviewed in subsection \ref{ssec:CLUSTERcft}.

A priori, however, there is no guarantee that such orbifold actions correspond
to endpoint theories. Rather, they could involve supplementing an endpoint
theory by additional blowups or higher singularities in the elliptic fiber, as
well as non-isolated singularities (non-compact flavor seven-branes).
Additionally, as we already saw in subsection \ref{ssec:CLUSTERcft}, not all
endpoint theories have such a simple elliptic fibration. Indeed, not even all rigid theories can be
represented as orbifolds. For example $\mathcal{C}_{end} = 92$ involves
a $\mathbb{Z}_{17}$ orbifold in the base.

\subsection{Orbifold Examples: D-type theories}
The group actions and resolutions for D-type theories can be found for example 
in reference \cite{UTWOorb}. Here, we focus on the value of the ratio $p / q$, with notation as in 
equation (\ref{Dorbo}).

The rigid endpoint $D_N24$ corresponds to the fraction
\begin{equation}
\frac pq = \frac{3}{3N-2}
\end{equation}
while the rigid endpoint $D_N32$ corresponds to the fraction
\begin{equation}
\frac pq = \frac{3}{3N-4}.
\end{equation}
We have also calculated the fraction for the endpoint $D_N23$:
\begin{equation}
\frac pq = \frac{2}{2N-1}.
\end{equation}

\section{Supplementing a Minimal Model \label{sec:SUPP}}

Our focus up to this point has been on a classification of endpoints. In this
section we discuss the various ways to go beyond these minimal cases. Recall
that starting from an endpoint, there is a unique way to perform a minimal
resolution, and there is also a minimum order of vanishing for $f$, $g$ and
$\Delta$ for the Weierstrass model. With enough foresight, instead of
specifying a minimal resolution, one could instead simply list the specific
order of vanishing for $f$, $g$ and $\Delta$ on each of the curves of an
endpoint configuration.

Now, it is sometimes possible to make the Kodaira-Tate fiber more singular on
a given curve, i.e., by increasing the order of vanishing of $f,g$ and
$\Delta$ over a given curve. This should be viewed as giving another SCFT
which upon smoothing can flow back down to one of our endpoint configurations.
In the blown up phase, this smoothing can be interpreted as a Higgs mechanism
for the seven-brane gauge theory.

To unambiguously assign a CFT to a given F-theory model, we therefore need to
give a configuration of curves, and then also specify the gauge algebra over
the curves and intersection points.\footnote{Recently it has been found that a
singular compactification of F-theory can contain various ambiguities, which
are repaired by including additional T-brane data \cite{Anderson:2013rka} (see
also \cite{TBRANES, glueI, glueII}). The important point for our
considerations is that non-trivial T-brane data corresponds to moving onto the
Higgs branch of a putative 6D\ SCFT, i.e., it introduces a mass scale through
the presence of a seven-brane flux and non-trivial Higgs field profile.
\par
For our present purposes, such phenomena should be viewed as specifying a
Higgs branch for the moduli space of a six-dimensional theory which already
has a geometric description. This can in principle lead to a new flow, but the
IR fixed point is still characterized by the gauge algebra on the
configuration of curves. Additionally, we do not expect to reach any isolated
6D SCFTs which are completely disconnected from a geometric realization.
\par
In four-dimensional SCFTs, T-brane data instead shows up as a mass deformation
of the CFT realized on the worldvolume of a D3-brane probing a stack of
seven-branes (see e.g. \cite{Funparticles, FCFT, HVW, D3gen}).} This sort of
supplementing leads us to new $(1,0)$ theories, and can be done for all of the
ADE $(2,0)$ theories, as well as many of the minimal $(1,0)$ SCFTs found in
section \ref{sec:LIST}.

As we just mentioned, one way to supplement a theory involves decorating the
curves of an endpoint configuration by a gauge algebra of higher rank than the
minimal one required to define an elliptic model. For example, in the A-type
$(2,0)$ theory given by a configuration of $-2$ curves, we can either leave
the curves undecorated, or make the elliptic fiber more singular by decorating
each curve by the same gauge algebra:\footnote{Here we omit non-compact divisors associated with flavor branes. Such contributions 
must be included to have an anomaly free theory.}
\begin{equation}
\mathcal{C}_{\text{generic}}=\underset{N}{\underbrace{%
\begin{tabular}
[c]{ccc}%
$2$ & $...$ & $2$%
\end{tabular}
}}\text{ \ \ versus \ \ }\mathcal{C}_{\text{tuned}}=\underset{N}{\underbrace{%
\begin{tabular}
[c]{ccc}%
$\mathfrak{su}(m)$ & $...$ & $\mathfrak{su}(m)$\\
$2$ & $...$ & $2$%
\end{tabular}
}},
\end{equation}
where by abuse of notation, we refer to the singularity type according to the
gauge symmetry it would generate in the resolved phase. This can be done for
any of the $(2,0)$ ADE theories, and gives a purely geometric realization of
the $(1,0)$ theories of \cite{Blum:1997mm} obtained from five-branes probing
an ADE singularity. Note that if the Kodaira-Tate fiber is more singular than
the generic model with a given base, then the gluing condition will not
involve a product algebra inside of $\mathfrak{e}_{8}$. Rather, it will
involve a product contained in a parent simple gauge algebra. For example, a
$-1$ curve can meet two curves carrying $\mathfrak{so}(m)$ and $\mathfrak{so}%
(n)$ gauge symmetry for arbitrary values of $m$ and $n$, and $\mathfrak{so}%
(m)\oplus\mathfrak{so}(n)\subset\mathfrak{so}(m+n)$
\cite{MR690264,Bershadsky:1996nu}. In general, once we go above the minimal
singular behavior for an elliptic fibration, there are additional consistency
conditions for gluing besides embedding a product in a simple parent algebra.

It is also sometimes possible to introduce other singular fibers, such as the
$II^{\ast}$ fiber associated with an $\mathfrak{e}_{8}$ gauge algebra. In such
cases, additional blowups in the base are generically required. A class of
examples covered in this way includes small instantons of heterotic theory on
top of an ADE\ singularity of a K3 surface \cite{Aspinwall:1997ye}. Under
heterotic/F-theory duality, the ADE\ singularity of the heterotic theory
becomes a stack of strongly coupled seven-branes with ADE\ gauge group.
Introducing additional pointlike instantons corresponds to further blowups of
the base geometry. Performing such blowups, we generically make the
singularity type worse, which in turn can require even more blowups.

Specifying a non-minimal theory amounts to introducing extra blowups in the
base, and decorating it with a consistent choice of gauge
algebras.\footnote{Note that decorating by the choice of gauge algebra rather
than just the Kodaira-Tate fiber allows us to also cover T-brane deformations
of the F-theory model.} These two steps are typically interrelated. First, we
can enlarge the class of possible bases by introducing non-minimal blowups of
an endpoint. Physically, this corresponds to incorporating additional E-string
theories. For each such base, there is a minimal order of vanishing for $f$,
$g$, and $\Delta$ for each contractible curve, which can be shifted up by
introducing extra blowups. Additionally, for a given base, we can decorate
each curve with a higher rank gauge algebra. This will in general violate a
gluing/gauging condition, so further blowups in the base will then be
required. Once we go beyond the minimal Kodaira-Tate type for the elliptic
fiber, there are also extra compatibility conditions besides just ``fitting
inside a parent simple gauge algebra''. It would be interesting to work out
these conditions in full generality. Finally, activating a deformation by
giving a vev to an operator in the SCFT can then induce flows to lower theories.

Let us illustrate how blowing up can shift the minimal singularity type of the
elliptic fibration. The most trivial example is to start with F-theory on the
base $B=\mathbb{C}^{2}$. Blowing up once, we get a single $-1$ curve, which is
the E-string theory. Blowing up again, we get the configuration $\mathcal{C}%
=12$. Blowing up again, we can either reach the theory $122$ or $131$, or
$213$. If we now blow down the $-1$ curve in the last two theories, we get a
theory $\mathcal{C}=12$, but where the $-2$ curves supports a gauge algebra.
This can be deformed down to the generic case where the $-2$ curve has a
smooth elliptic fiber.

As a more involved example, consider next the endpoint configuration and its
minimal resolution:
\begin{equation}
\mathcal{C}_{\text{end}}=33,\,\,\,\text{and}\,\,\,\mathcal{C}_{\text{min}%
}=414=%
\begin{tabular}
[c]{ccc}%
$\mathfrak{so}(8)$ &  & $\mathfrak{so}(8)$\\
$4$ & $1$ & $4$%
\end{tabular}
\ ,
\end{equation}
where we have indicated the minimal singularity type. Now, we can choose to
perform some further blowups, and doing so leads us to a new theory. For
example, an additional blowup on the $-4$ curve on the right takes us to:%
\begin{equation}
\mathcal{C}_{\text{n-min}}\rightarrow4151.
\end{equation}
We cannot stop here, however, because our non-Higgsable clusters now violate
the local gauging condition for nearest neighbors! We therefore need to go
through the same steps we did for the endpoints, now for this non-minimal
case. The process terminates after one further blowup:%
\begin{equation}
\mathcal{C}_{\text{min}}\rightarrow513161=%
\begin{tabular}
[c]{cccccc}%
$\mathfrak{f}_{4}$ &  & $\mathfrak{su}(3)$ &  & $\mathfrak{e}_{6}$ & \\
$5$ & $1$ & $3$ & $1$ & $6$ & $1$%
\end{tabular}
\ ,
\end{equation}
which satisfies the gauging condition. Observe that by a further tuning, we
can supplement $\mathfrak{f}_{4}$ to convert it to an $\mathfrak{e}_{6}$
singularity type (i.e., no monodromy in the $IV^{\ast}$ fiber):
\begin{equation}
\mathcal{C}_{\text{tuned}}=%
\begin{tabular}
[c]{cccccc}%
$\mathfrak{e}_{6}$ &  & $\mathfrak{su}(3)$ &  & $\mathfrak{e}_{6}$ & \\
$5$ & $1$ & $3$ & $1$ & $6$ & $1$%
\end{tabular}
\ .
\end{equation}
This still satisfies the gauging condition, so no further blowups would be
required in that case either.

Of course, there is no need to stop at a single additional blowup. Adding
another blowup on the same configuration, we can instead take:%
\begin{equation}
\mathcal{C}_{\text{nn-min}}=%
\begin{tabular}
[c]{|l|l|l|}\hline
& $1$ & \\\hline
$41$ & $6$ & $1$\\\hline
\end{tabular}
\ \ \ .
\end{equation}
Going through the blowups of the base in this case, we have:%
\begin{equation}
\mathcal{C}_{\text{nn-min}}\rightarrow%
\begin{tabular}
[c]{|l|l|l|}\hline
& $1$ & \\\hline
$5131$ & $7$ & $1$\\\hline
\end{tabular}
\rightarrow%
\begin{tabular}
[c]{|l|l|l|}\hline
& $1$ & \\\hline
$51321$ & $8$ & $1$\\\hline
\end{tabular}
\end{equation}
There are clearly many possible ways of introducing additional non-minimal
blowups. To avoid clutter, we have suppressed the minimal gauge algebras for
these examples.

Some (but not all!) endpoint configurations admit an infinite sequence of such
blowups. Consider for example the endpoint $\mathcal{C}_{\text{end}}=3$.
Adding successive blowups to the rightmost curve in the graph, we reach, after
$k$ blowups, a configuration:
\begin{equation}
\mathcal{C}_{\text{n-min}}=4\underset{k}{\underbrace{2...21}}.
\end{equation}
Now, this admits a consistent resolution to non-Higgsable clusters because the
configuration $42...2$ is an endpoint configuration with \textquotedblleft
room to spare\textquotedblright\ in the minimal resolution.

Many theories only admit a finite number of additional blowups. This appears
to be the generic situation when there are a sufficiently large number of $-2$
curves in an endpoint. To give an example of this type, consider the endpoint
given by the $E_{7}$ graph:%
\begin{equation}
\mathcal{C}_{E_{7}}=%
\begin{tabular}
[c]{|l|l|l|l|l|l|}\hline
&  & $2$ &  &  & \\\hline
$2$ & $2$ & $2$ & $2$ & $2$ & $2$\\\hline
\end{tabular}
\ \ \ .
\end{equation}
Blowing up any point eventually requires us to blowup the $-2$ curve which
only touches the trivalent vertex. So, consider first blowing up at this
point. There is then a triggered sequence of blowups, leading eventually to:%
\begin{equation}
\mathcal{C}_{E_{7}}\rightarrow...\rightarrow%
\begin{tabular}
[c]{|l|l|l|l|l|l|l|l|}\hline
&  &  & $3$ &  &  &  & \\\hline
&  &  & $1$ &  &  &  & \\\hline
$2$ & $3$ & $1$ & $5$ & $1$ & $3$ & $2$ & $2$\\\hline
\end{tabular}
\ \ \ .
\end{equation}
Anywhere else we attempt to blowup will eventually force more than seven
blowups on the $-5$ curve. So this theory only admits one extra blowup. From the
structure of the NHCs, we can also read off the gauge symmetry for this theory:
\begin{equation}
\begin{tabular}
[c]{|l|l|l|l|l|l|l|l|}\hline
&  &  & $\mathfrak{su}(3)$ &  &  &  & \\\hline
&  &  & $\oplus $ &  &  &  & \\\hline
$\mathfrak{su}(2)$ & $\mathfrak{g}_2$ & $\oplus$ & $\mathfrak{f}_4$ & $\oplus$ & $\mathfrak{g}_2$ & $\mathfrak{sp}(1)$ & \\\hline
\end{tabular}
\end{equation}
This can be viewed as a gauging of the $\mathfrak{e}_{8}^{\oplus 3}$
flavor symmetries for the three E-string theories associated with the $-1$ curves, according to the
connectivity of the diagram. For example, the $\mathfrak{f}_4$ factor gauges the diagonal subalgebra of
$\mathfrak{e}_{8}^{\oplus 3}$.

There are also minimal theories which are completely rigid. An example of this
type is the endpoint $\mathcal{C}_{\text{end}}=(12)$, i.e., a single $-12$
curve. Any further blowups would make the self-intersection too high, and the
Kodaira-Tate type of the elliptic fiber is already maximally singular. Another
example is the $E_{8}$ endpoint with graph:%
\begin{equation}
\mathcal{C}_{E_{8}}=%
\begin{tabular}
[c]{|l|l|l|l|l|l|l|}\hline
&  & $2$ &  &  &  & \\\hline
$2$ & $2$ & $2$ & $2$ & $2$ & $2$ & $2$\\\hline
\end{tabular}
\ \ \ .
\end{equation}
If we attempt to blowup anywhere, it will eventually involve the $-2$ curve
which only touches the trivalent vertex. However, the sequence of forced
blowups leads us to:%
\begin{equation}
\mathcal{C}_{E_{8}}\rightarrow...\rightarrow%
\begin{tabular}
[c]{|l|l|l|l|l|l|l|l|l|l|l|l|l|l|l|l|}\hline
&  &  &  &  & $3$ &  &  &  &  &  &  &  &  &  & \\\hline
&  &  &  &  & $2$ &  &  &  &  &  &  &  &  &  & \\\hline
&  &  &  &  & $2$ &  &  &  &  &  &  &  &  &  & \\\hline
&  &  &  &  & $1$ &  &  &  &  &  &  &  &  &  & \\\hline
$2$ & $3$ & $2$ & $2$ & $1$ & $(12)$ & $1$ & $2$ & $2$ & $3$ & $1$ & $5$ & $1$
& $3$ & $2$ & $2$\\\hline
\end{tabular}
\ \ \ ,
\end{equation}
which still needs a further blowup on the left leg of the graph. This in turn
pushes the $-12$ curve beyond the saturation limit, so no blowups can be
tolerated for this theory. Finally, a more involved class of such rigid
theories is:%
\begin{equation}
\mathcal{C}_{7A_{N}7}=7\underset{N}{\underbrace{2...2}}7.
\end{equation}
Performing the minimal resolution, the self-intersection of the $-7$ curves
shifts to $-12$, and the self-intersection of the $-2$ curves also shifts to
$-12$. Any further blowups would lead to an inconsistent model.

\section{Flowing Through Blowdowns \label{sec:FLOWS}}

So far, our approach to classification has involved a two stage process.
First, we give a choice of minimal base and singular elliptic fibration, and
then proceed to supplement this class of theories. Supplementing involves possibly further
blowups in the base, as well as decoration of all curves by a consistent choice of gauge algebra.
In this way, we have given a systematic way of generating all possible SCFTs which can be realized in
F-theory. This should be viewed as a \textquotedblleft bottom
up\textquotedblright\ approach to the classification of 6D SCFTs.

Given such a list of SCFTs, it is natural to ask how such theories are connected by flows. We defer a more complete analysis
to future work \cite{BackToTheFuture}, and so will instead focus on the physical interpretation of blowdowns in the base. Recall that these
blowdowns involve shrinking down various $-1$ curves, and then possibly performing a move in the complex structure moduli. It is therefore
natural to expect that this corresponds to a flow in the SCFT moduli space.

The way we establish the possibility of a flow by such blowdowns
is somewhat technical, but we include it here for completeness.
We begin with a base $B$ corresponding to one of our endpoints, which takes
the form $\mathbb{C}^{2}/\Gamma$. There is a one-dimensional representation
$V_{m}$ of $\Gamma$ corresponding to $-mK_{B}$, and we can build appropriate
spaces\footnote{These spaces do not always form bundles, but can be treated as
coherent sheaves if the need arises.} $(\mathbb{C}^{2}\times V_{m})/\Gamma$ in
which $x,y,f,g$ take values (for $m=2,3,4,6$, respectively), and define an
elliptic fibration by the familiar Weierstrass equation $y^{2}=x^{3}+fx+g$. We
include the case of trivial $\Gamma$ in our considerations.

Because $B$ is non-compact, there is no limit on the degrees of monomials in
the $\mathbb{C}^{2}$ coordinates which can occur in $f$ and $g$. However, for
any specific SCFT, the degrees of monomials which can affect that SCFT are
bounded. We will work with $f$ and $g$ polynomials of some fixed large degree,
keeping in mind that  we can enlarge the allowed degree if necessary.

For each pair $(f,g)$, there is either a specific blowup of $B$ uniquely
determined by the pair over which $(f,g)$ produce a nonsingular F-theory
model, or $(f,g)$ are too singular to produce a nonsingular F-theory model on
any blowup. We will now describe how to determine the blowup, and whether it exists.

We begin with the blowup $B_{k}\to B$ which is the minimal resolution of
singularities of the orbifold. We use $k$ to label the number of compact curves
on $B_{k}$. The Weierstrass coefficients $f$ and $g$ pull back to sections
$f_{k}$, $g_{k}$ of appropriate line bundles on $B_{k}$. There are three
possibilities. First, if $f_{k}$ vanishes to order at least $4$ and $g_{k}$
vanishes to order at least $6$ on any (compact) curve on $B_{k}$, then we
discard $(f,g)$ as these coefficients  cannot lead to a nonsingular F-theory
model on any blowup of $B_{k}$.

Second, if at every point on $B_{k}$, either the multiplicity of $f_{k}$ at
the point is at most $3$ or the multiplicity of $g_{k}$ at the point is at most
$5$, then we do not need any further blowups: $(f_{k},g_{k})$ define a
nonsingular F-theory model over $B_{k}$ itself.

Third, if neither of the first two cases hold, then there are finitely many
points on $B_{k}$ at which $f_{k}$ has multiplicity at least $4$ and $g_{k}$
has multiplicity at least $6$; pick one such point and blow it up, to obtain a
new surface $B_{k+1}$ and a blowdown map $\rho:B_{k+1}\to B_{k}$. We define
$f_{k+1}=\rho^{*}(f_{k})/\sigma^{4}$ and $g_{k+1}=\rho^{*}(g_{k})/\sigma^{6}$,
where $\sigma=0$ is a defining equation for the $-1$ curve created by the
blowup. Then $(f_{k+1},g_{k+1})$ are sections of the appropriate bundles to be
the Weierstrass coefficients of an F-theory model over $B_{k+1}$.

We now repeat this process for $B_{k+1}$, obtaining the same three
alternatives as before. Since our polynomials have only a finite number of
terms, and each of those multiplicity requirements imposes linear conditions
on the terms in the polynomials, for any specific $(f,g)$ this process must
stop after a finite number of steps. At the end of the process, either the
pair $(f,g)$ has been eliminated due to excessive vanishing along some curve,
or a blowup $B_{N}$ (with $N$ curves) has been obtained over which
$(f_{N},g_{N})$ define a nonsingular F-theory model. Note that no further
blowups are allowed with this particular $f$ and $g$.

In the blown up phase, the F-theory model defined by $(f_{N},g_{N})$ has a
number of characteristics: there are $N$ tensor multiplets, gauge multiplets
which are determined by the Kodaira-Tate singularity types along the various
curves in the configuration, charged matter multiplets which are determined by
the way the curves in the configuration meet the discriminant locus, and
neutral matter multiplets which are determined by the polynomials $f_{N}$ and
$g_{N}$. When we shrink the configuration of curves on $B_{N}$ to obtain a
superconformal fixed point, most of the higher degree terms in $f_{N}$ and
$g_{N}$ will not affect the actual fixed point we reach: They are associated
with irrelevant deformations of the SCFT. The key point is that the leading
order behavior of the singular fiber still controls the choice of SCFT we reach.

In particular, even though there is no Lagrangian description available, the
singularity of the elliptic fiber remains well-defined, so there is still a
notion of the \textquotedblleft gauge algebra\textquotedblright\ associated
with the singular model. We saw examples of this type in section
\ref{sec:SUPP} where we could decorate a minimal endpoint configuration by a
higher order singularity type.

As explained in section \ref{sec:SUPP}, to each blown up nonsingular F-theory model,
then, we can associate a graph describing the configuration of curves, and we can decorate
each node of this graph with a gauge algebra consistent
with a local gauging/gluing condition. To reach the
superconformal fixed point, we then pass to the strongly
coupled regime of this theory by shrinking the curves in the base to zero
size. Note that since the elliptic model is governed by a complex
equation, it is still possible to track the behavior of the geometry in
this limit.

In mathematical terms, the restriction on these new labels is that each curve
must be labeled with a singularity which is at least as singular as the
minimal singularity for that curve (i.e., the singularity in a maximally
Higgsed model with the same curve configuration). Moreover, the singularities
on two curves which meet cannot be so bad that they force a point of
multiplicities $(4,6)$ in $f$ and $g$ at their intersection. The singularity
types are characterized by orders of vanishing of $f_{N}$ and $g_{N}$ as well
as some data about whether certain expressions built from $f_{N}$ and $g_{N}$
have single-valued square roots (i.e., monodromy in the fiber). For further
discussion see for example \cite{Grassi:2011hq}.

As an example, consider blowing down the
$-1$ curve in $213$ to reach a $12$ theory with a decorated $-2$ curve.
We can summarize this by the flows:
\begin{equation}%
\begin{tabular}
[c]{ccc}
&  & $\mathfrak{su}(3)$\\
$2$ & $1$ & $3$%
\end{tabular}
\overset{RG}{\longrightarrow}%
\begin{tabular}
[c]{cc}
& $\mathfrak{su}(3)$\\
$1$ & $2$%
\end{tabular}
\overset{RG}{\longrightarrow}%
\begin{tabular}
[c]{cc}
& \\
$1$ & $2$%
\end{tabular}
\end{equation}

Proceeding in this way, we can also see that the generic configuration $122$
defines a different SCFT. To see this, observe that upon blowdown of a $-1$
curve in the configuration $213$, we reach:%
\begin{equation}
\mathcal{C}_{\text{tuned}}=%
\begin{tabular}
[c]{cc}
& $\mathfrak{su}(3)$\\
$1$ & $2$%
\end{tabular}
,
\end{equation}
i.e., a $-2$ curve decorated by a singular Kodaira-Tate fiber. This is to be
contrasted with the blowdown of $122$ to $\mathcal{C}_{\text{generic}}=12$
with no singular elliptic fiber. The conclusion is that there is a further
relevant deformation/unfolding from $\mathcal{C}_{\text{tuned}}$ to
$\mathcal{C}_{\text{generic}}$. This is a relevant deformation, because it
affects the leading order behavior of the singularity type.

\section{Conclusions \label{sec:CONC}}

In this paper we have studied geometrically realized superconformal field
theories which can arise in compactifications of F-theory. Our primary claim
is that minimal 6D SCFTs are captured by F-theory on the
non-compact base $\mathbb{C}^{2}/\Gamma$ with $\Gamma$ a discrete subgroup of
$U(2)$. Furthermore, one can obtain additional non-minimal SCFTs in two
possible ways: first, by bringing in a number of additional E-string
theories and second, by decorating a configuration of curves with a non-minimal gauge algebra.
Both these operations lead to SCFTs with more degrees of freedom that can be deformed to
flow down to lower theories, including all of the minimal ones. In
the rest of this section we discuss some avenues of future investigation.

It is natural to ask whether there are additional SCFTs in six dimensions. Our
list includes all the known examples of SCFTs in six dimensions that we are
aware of, as well as a vast number of previously unknown examples. Based on
this, it is tempting to conjecture that our method generates
the full list of 6D SCFTs. It would be interesting to see if this is true.

We have fully classified the minimal 6D SCFTs and have indicated how to go
about classifying non-minimal ones by introducing further blowups in the base, and decorating a
graph of curves by a non-minimal gauge algebra. It would be quite instructive to
give an explicit characterization of this class as well.

Regardless of whether there are even more 6D SCFTs outside our list, it should be clear that
we have established the existence of a vast number of previously unknown $(1,0)$ theories.
This should provide an excellent theoretical laboratory for gaining further insight into the workings
of conformal theories in six dimensions.

For example, in this paper we have mainly focussed on the singular limit associated with a
given conformal fixed point. To gain further insight into the structure of these theories,
it would be exciting to fully map out the various branches of moduli space. We expect that
this should be possible using the geometry of the F-theory model.

It would also be interesting to study compactifications of our $(1,0)$
theories to lower dimensions and determine which ones flow to an interacting
lower-dimensional SCFT. For the case of $E$-string theories this has already been
studied and leads to interesting theories in lower dimensions. It is reasonable
to believe that this may also be the generic story for many of the additional
$(1,0)$ SCFTs we have discovered in six dimensions.

Finally, we have discovered several new infinite families of $(1,0)$ SCFTs
which are labeled by an integer $N$ which can be taken arbitrarily large. It
would be interesting to determine whether there is a smooth large $N$
limit of these theories, and a corresponding gravity dual, perhaps along the lines
of \cite{Apruzzi:2013yva}.

\section*{Acknowledgements}

We thank M. Del Zotto, B. Haghighat, T. Dumitrescu, D.S. Park and E. Witten for helpful discussions. We
also thank the 2013 Summer Workshop at the Simons Center for Geometry and
Physics for hospitality, where some of this work was completed. The work of
JJH and CV is supported by NSF grant PHY-1067976. The work of DRM\ is
supported by NSF grant PHY-1307513.



\appendix

\section{Instructions for Using the Mathematica Notebook \label{APP:mathematica}}

In the \texttt{arXiv} submission we also include a sample \texttt{Mathematica} notebook called \texttt{BlowUpDown.nb}.
This file computes the minimal resolution of a candidate endpoint configuration, and also determines
whether a given linear chain of NHCs blows down to a consistent endpoint.

To access the \texttt{Mathematica} notebook, proceed to the URL where the \texttt{arXiv} submission
and abstract is displayed. On the righthand side of the webpage, there will be
a box labeled ``Download:''. Click on the link ``Other formats'', and
then under the boldfaced line ``Source'' click on the link
``[ Download source ]''. A download of a zipped file should then commence. In some cases, it may be necessary to append the ending .tar.gz to the
end of the file. The set of submission files along with the \texttt{Mathematica} notebook can then be accessed by unzipping this file. For further
instruction on unzipping such files, see for example \texttt{http://arxiv.org/help/unpack}.

The two main functions in the file are \texttt{BlowUP} and \texttt{BlowDOWN} which take as input a vector of integers
$\{a_1,...,a_r \}$. The respective outputs are a computation of the minimal resolution, and sequential
blowdown of all $-1$ curves. The functions also indicate if a configuration of curves contains various inconsistencies. We also include
an example of how to perform a sweep over all configurations for ten or fewer curves in an A-type graph. We have not optimized the notebook scripts for
efficiency, but even so, on a (circa 2013)
laptop the full computation only takes about half a day to finish.

\section{Constraints on Contractible Curve Configurations} \label{APP:constraints}

Suppose that $\Sigma_1$, \dots, $\Sigma_r$ are compact complex curves on a
complex surface $B$.
Mumford \cite{MR0153682} showed that if those curves can contract to
a point (smooth or singular), then the negative of the intersection
matrix
\begin{equation}
A_{ij}=-(\Sigma_i\cdot\Sigma_j)
\end{equation}
is positive definite.
  Conversely, if that matrix is positive definite,
then Grauert \cite{MR0137127} showed that the collection of curves can
be contracted to a complex analytic (singular) point.  Artin \cite{MR0146182}
showed that if, in addition, the singularity satisfies a condition known
as ``rationality'' \cite{MR0199191}, the contraction is guaranteed to
be an algebraic singular point.  In fact, all orbifolds are rational
singularities, so when we contract we stay within the realm of algebraic
varieties.

Given any divisor $D$ on $B$, there are nonnegative rational numbers
$a_i$ such that
\begin{equation}
\left(D-\sum a_i\Sigma_i\right)\cdot \Sigma_j
\ge0 \text{ for all $j$,}
\end{equation}
and
\begin{equation}
\left(D-\sum a_i\Sigma_i\right)\cdot \Sigma_j
= 0 \text{ whenever } a_j\ne0.
\end{equation}
The $\mathbb{Q}$-divisor $N=\sum a_i\Sigma_i$ is called the
{\em negative part of the Zariski decomposition of $D$} \cite{MR0141668}.
The coefficients $a_i$ depend on the divisor $D$.

Let $\sum \alpha_i\Sigma_i$ be the negative part of the Zariski decomposition
of $-K_B$.  Then any section of $-mK_B$ must vanish to order at least
$m\alpha_i$ along $C_i$.  Since $m\alpha_i$ is in general a rational number,
what this means in practice is that the order of vanishing must be the
next highest integer (as a minimum).

From this we infer an important property:  if $\alpha_i>5/6$, then
$f$ and $g$  vanish along $\Sigma_i$ to orders $4$ and $6$, respectively.
  In particular, a base containing
such a configuration cannot support a nonsingular
F-theory model, even after blowing up the base further.  To see this, notice that
$4\alpha_i>10/3$ so the minimum order of vanishing of $f$ is $4$, and
$6\alpha_i>5$ so the minimum order of vanishing of $g$ is $6$.

Now we use this property to formulate some constraints on contractible
curve configurations.  First, every curve $\Sigma$ in the configuration
must have arithmetic genus $g=0$.  For if $\Sigma$ were a curve with
positive arithmetic genus, then $K_B\cdot \Sigma + \Sigma\cdot \Sigma
=2g-2\ge0$.  It follows that
\begin{equation}
1+\frac{2-2g}{\Sigma\cdot\Sigma}\ge1>\frac56\text{ and }
\left(-K_B-\left(1+\frac{2-2g}{\Sigma\cdot\Sigma}\right)\Sigma\right)\cdot \Sigma=0.
\end{equation}
Thus, the coefficient of $\Sigma$ in the negative part of the Zariski
decomposition of $-K_B$ is greater than $5/6$, so there is no nonsingular
F-theory model supported by this surface.

Second, given two curves $\Sigma$ and $\Sigma'$, they can meet in at
most one point.  For if $\Sigma\cdot\Sigma'\ge2$, then
\begin{equation}
\left(-K_B-\Sigma-\Sigma'\right)\cdot \Sigma =2-\Sigma'\cdot\Sigma\le0,
\end{equation}
and similarly $(-K_B-\Sigma-\Sigma')\cdot\Sigma'\le0$.  Let $a\Sigma+a'\Sigma'$
be the negative part of the Zariski decomposition of $-K_B-\Sigma-\Sigma'$.
Then $(1+a)\Sigma+(1+a')\Sigma'$ is the negative part of the Zariski
decomposition of $-K_B$.  Since $1+a$ and $1+a'$ are both greater than $5/6$,
$f$ and $g$ vanish to too high an order along $\Sigma$ and $\Sigma'$
so there is no nonsingular F-theory model.

Finally, let $\Sigma_1$, \dots, $\Sigma_r$ be a loop (of curves of genus
$0$).  Then
\begin{equation}
\left(-K_B-\sum_i \Sigma_i\right)\cdot \Sigma_j =
-K_B\cdot\Sigma_j-\Sigma_j\cdot\Sigma_j - \left(\sum_{i\ne j}\Sigma_i\right)\cdot\Sigma_j
=2-2=0,
\end{equation}
since each curve in the loop meets exactly two other curves in the loop.
Again, this shows that the coefficients in the negative part of the Zariski
decomposition of $-K_B$ are all $1$, and in particular all greater than
$5/6$, so there is no nonsingular F-theory model.

As a special case, notice that three curves passing through a common
point have the same intersection properties as three curves in a loop,
so this is excluded as well.

The conclusion is that, in order to support a nonsingular F-theory model (even
after blowing the base up further),
our collection of curves must all be rational, meeting two at a time
in distinct points and
with no tangencies (i.e., it has ``normal crossings''), and the dual
graph of the collection must form a tree.

\section{The Gluing Condition in the Maximally Higgsed Case} \label{APP:gluing}

Let $B$ be a neighborhood of a contractible configuration of compact
curves which supports a nonsingular F-theory model, and let $E$ be
a $-1$ curve on $B$.  We choose a nonsingular F-theory model which is
maximally Higgsed, and consider the other curves $\Sigma_j$
in the configuration  which meet $E$.  The ``gluing condition'' which
we would like to verify states that the direct sum of the Lie algebras
$\mathfrak{g}_{\Sigma_j}$ must be a subalgebra of $\mathfrak{e}_8$.

Verifying this condition means analyzing all of the ways in which the
other curves can meet $E$.  In fact, table 3 of
\cite{Morrison:2012np} already listed all ways that $-1$ curves
can meet non-Higgsable clusters, but here we are much more constrained:
because our $-1$ curves are part of a contractible configuration of
curves, and the configuration must have normal crossings even after
blowing down the $-1$ curve, each $(-1)$ curve can meet at most
two other curves (and it will meet them with intersection number $1$).

Since each of the gauge algebras which occurs in a non-Higgsable cluster
is a subalgebra of $\mathfrak{e}_8$, the condition is automatically
satisfied if $E$ meets $0$ or $1$ curves.  In the case of $2$ curves,
blowing down $E$ gives us two components of the discriminant locus which
meet each other (or ``collide''):  this situation was analyzed and
applied to F-theory long ago \cite{MR690264,Bershadsky:1996nu}.  However,
those works had an additional hypothesis (``well-defined $J$ invariant'')
which is not relevant here, so we need to do the analysis again.

Before beginning, we recall a refinement of the classification
of clusters from \cite{Morrison:2012np}.  Each curve in each
cluster not only carries a gauge algebra, but a specific Kodaira-Tate type,
labeled by the orders of vanishing of $f$ and $g$ along the curve.
This even applies to the curve in the $3,2,2$ cluster which carries no
gauge algebra:  it has a Kodaira type labeled by a single order of
vanishing of both $f$ and $g$; we will label this in parallel
to the other gauge algebras as $\mathfrak{h}_0$ (following the notation of
\cite{Ganor:1996pc}.)\footnote{In fact, as pointed out in \cite{Ganor:1996pc},
when compactified to four dimensions, this Kodaira type leads to
the Argyres-Douglas SCFT \cite{Argyres:1995jj}.}
The gauge algebra $\mathfrak{su}(2)$ actually has two realizations
from this point of view; we refer to one of them as $\spone$ in order
to keep them straight.

Here is a list of all possible gauge algebras, together with their
Kodaira conditions (where $\mathfrak{h}_0$ is trivial as an algebra).
\begin{equation}
\begin{tabular}{c|ccccccccccc}
algebra & $\mathfrak{h}_0$ & $\sutwo$ &
$\spone$ &$\mathfrak{su}(3)$
&$\mathfrak{g}_2$&$\mathfrak{so}(7)$&$\mathfrak{so}(8)$&
$\mathfrak{f}_4$&$\mathfrak{e}_6$&$\mathfrak{e}_7$&$\mathfrak{e}_8$\\ \hline
$\ord_\Sigma(f)$ & 1&1& 2 & 2 &2&2&2&3&3&3&4\\
$\ord_\Sigma(g)$ & 1&2&2 &2&3&3&3&4&4&5&5\\
\end{tabular}
\end{equation}
The $3,2$ cluster has algebras
$\mathfrak{g}_2, \sutwo$, the $3,2,2$
cluster has algebras $\mathfrak{g}_2, \spone, \mathfrak{h}_0$,
and the $2,3,2$ cluster has algebras
$\sutwo, \mathfrak{so}(7), \sutwo$.

Our next observation is that a curve with algebra $\spone$ cannot
be one of two curves meeting $E$.  This is because the only curve
carrying $\spone$ is the central curve in a $3,2,2$ cluster.  If it
was one of two curves meeting $E$, then as shown in figure~\ref{no-sp1},
after two blowdowns we lose normal crossings, which is not permitted.
Thus, whenever we see
orders of vanishing $(2,2)$ for $(f,g)$, we may assume that the gauge
algebra is $\mathfrak{su}(3)$.

\begin{figure}[ptb]
\begin{center}
\includegraphics{fig2a.mps}
\qquad
\includegraphics{fig2b.mps}
\qquad
\raisebox{18pt}{\includegraphics{fig2c.mps}}
\end{center}
\caption{Blowing down a configuration containing $\spone$}
\label{no-sp1}
\end{figure}

Now our basic computation is to consider a chain of three curves
$\Sigma_L$, $E$, $\Sigma_R$ and specify the orders of vanishing
of $f$ and $g$ along them as $(a_L,b_L)$, $(a,b)$, and $(a_R,b_R)$.
If we contract $E$ we will find a point where the multiplicities of
$f$ and $g$ are at least $a_L+a_R$ and $b_L+b_R$, respectively.  Therefore,
$a\ge a_L+a_R-4$ and $b\ge b_L+b_R-6$.  Let $a_{\text{min}}$ be the larger
of $a_L+a_R-4$ and $0$, and similarly let $b_{\text{min}}$
be the larger of $b_L+b_R-6$ and $0$.

Now, if $a_{\text{min}}\ge1$ and $b_{\text{min}}\ge2$ there will be a gauge algebra associated to
$E$, contrary to our assumption that we are working with a maximally
Higgsed F-theory model.  Thus, there are upper bounds on $a_{\text{min}}$ and $b_{\text{min}}$
and it is straightforward to make a table of allowed possibilities.

We give the table in two parts:  first, there are some cases which
do not actually occur:
\begin{equation}
\begin{tabular}{c|ccc|c}
$\mathfrak{g}_{\Sigma_L}$ & $(a_L,b_L)$ & $(a_{\text{min}},b_{\text{min}})$ & $(a_R,b_R)$ & $\mathfrak{g}_{\Sigma_R}$ \\ \hline
$\sutwo$ & $(1,2)$ &$(1,1)$ & $(4,5)$ & $\mathfrak{e}_8$\\
$\mathfrak{su}(3)$ & $(2,2)$ & $(1,1)$ & $(3,5)$ &
 $\mathfrak{e}_7$ \\ \hline
$\mathfrak{so}(7)$ or $\mathfrak{so}(8)$ &
  $(2,3)$ & $(1,1)$ & $(3,4)$ & $\mathfrak{f}_4$ or $\mathfrak{e}_6$ \\
$\mathfrak{g}_2$ or $\mathfrak{so}(7)$ or $\mathfrak{so}(8)$ &
  $(2,3)$ & $(1,1)$ & $(3,4)$ & $\mathfrak{e}_6$ \\
\end{tabular}
\end{equation}
The first two cases do not occur because the point of intersection of
$E$ and $\Sigma_R$ has multiplicities at least $(4,6)$ and thus it must
be blown up (contrary to our assumption that $B$ is the base of a
nonsingular F-theory model).  The last two cases do not occur because
of a interesting subtlety.  Although $E$ carries no gauge symmetry, it
is of Kodaira type $\mathfrak{h}_0$ in those cases, and in particular
$\{f=0\}$ must meet $\Sigma_L$ and $\{g=0\}$ must meet $\Sigma_R$.

On the other hand, if $\Sigma_R$ has gauge algebra $\mathfrak{e}_6$
then $(-6K_B-4\Sigma_R)\cdot \Sigma_R=-24+24=0$ so $\Sigma_R$ cannot
meet $\{g=0\}$.  If $\Sigma_L$ has gauge algebra $\mathfrak{so}(8)$
then $(-4K_B-2\Sigma_L)\cdot \Sigma_L = -8+8=0$ so $\Sigma_L$
cannot meet $\{f=0\}$.  And finally, if $\Sigma_L$ has gauge
algebra $\mathfrak{so}(7)$ then $\Sigma_L$ is the central curve is
a $2,3,2$ cluster, and for such a cluster, the effective divisor from
the Zariski decomposition is $-4K_B-\Sigma_L'-2\Sigma_L-\Sigma_L''$,
which has intersection number $0$ with $\Sigma_L$.  This again implies
that $\Sigma_L$ cannot meet $\{f=0\}$.

Having excluded those four cases, we are left with the two kinds of cases.
First, there are those in which both  $a_L+a_R-4$ and $b_L+b_R-6$
are nonnegative:
\begin{equation}
\begin{tabular}{c|ccc|c}
$\mathfrak{g}_{\Sigma_L}$ & $(a_L,b_L)$ & $(a_{\text{min}},b_{\text{min}})$ & $(a_R,b_R)$ & $\mathfrak{g}_{\Sigma_R}$ \\ \hline
$\mathfrak{h}_0$ & $(1,1)$ & $(0,0)$ & $(3,5)$ & $\mathfrak{e}_7$ \\
$\mathfrak{h}_0$ & $(1,1)$ & $(1,0)$ & $(4,5)$ & $\mathfrak{e}_8$ \\
$\sutwo$ & $(1,2)$ &$(0,0)$ & $(3,4)$ &
  $\mathfrak{f}_4$ or $\mathfrak{e}_6$ \\
$\sutwo$ & $(1,2)$ &$(0,1)$ & $(3,5)$ & $\mathfrak{e}_7$ \\
$\mathfrak{su}(3)$ & $(2,2)$ & $(1,0)$ & $(3,4)$ &
  $\mathfrak{f}_4$ or $\mathfrak{e}_6$ \\
$\mathfrak{g}_2$ or $\mathfrak{so}(7)$ or $\mathfrak{so}(8)$ &
  $(2,3)$ & $(0,0)$ & $(2,3)$ & $\mathfrak{g}_2$ or $\mathfrak{so}(7)$
  or $\mathfrak{so}(8)$\\
$\mathfrak{g}_2$ &   $(2,3)$ & $(1,1)$ & $(3,4)$ & $\mathfrak{f}_4$\\
\end{tabular} \label{eq:firstcases}
\end{equation}
Second, we have cases where either
$a_L+a_R-4$ or $b_L+b_R-6$ is negative; this
means that at least one additional zero of $f$ or $g$ occurs at the
intersection point, in addition to the ones coming from $\Sigma_L$ and
$\Sigma_R$.  These cases are:
\begin{equation}
\begin{tabular}{c|ccc|c}
$\mathfrak{g}_{\Sigma_L}$ & $(a_L,b_L)$ & $(a_{\text{min}},b_{\text{min}})$ & $(a_R,b_R)$ & $\mathfrak{g}_{\Sigma_R}$ \\ \hline
$\mathfrak{h}_0$ & $(1,1)$ & $(0,0)$ & $(1,1)$ & $\mathfrak{h}_0$ \\
$\mathfrak{h}_0$ & $(1,1)$ & $(1,0)$ & $(1,2)$ & $\sutwo$ \\
$\mathfrak{h}_0$ & $(1,1)$ &$(0,0)$ & $(2,2)$ &
  $\mathfrak{su}(3)$ \\
$\mathfrak{h}_0$ & $(1,1)$ &$(0,1)$ & $(2,3)$ & $\mathfrak{g}_2$ or $\mathfrak{so}(7)$ or $\mathfrak{so}(8)$ \\
$\mathfrak{h}_0$ & $(1,1)$ & $(1,0)$ & $(3,4)$ &
  $\mathfrak{f}_4$ or $\mathfrak{e}_6$ \\
$\sutwo$ & $(1,2)$ & $(0,0)$ & $(1,2)$ & $\sutwo$ \\
$\sutwo$ & $(1,2)$ & $(1,0)$ & $(2,2)$ & $\mathfrak{su}(3)$ \\
$\sutwo$ & $(1,2)$ &$(0,0)$ & $(2,3)$ &
$\mathfrak{g}_2$ or $\mathfrak{so}(7)$ or $\mathfrak{so}(8)$ \\
$\mathfrak{su}(3)$ & $(2,2)$ &$(0,1)$ & $(2,2)$ & $\mathfrak{su}(3)$ \\
$\mathfrak{su}(3)$ & $(2,2)$ & $(1,0)$ & $(2,3)$ &
$\mathfrak{g}_2$ or $\mathfrak{so}(7)$ or $\mathfrak{so}(8)$ \\
\end{tabular} \label{eq:secondcases}
\end{equation}
As can be easily seen, in each case listed in
(\ref{eq:firstcases}) and (\ref{eq:secondcases}), we have
$\mathfrak{g}_{\Sigma_L}\oplus \mathfrak{g}_{\Sigma_R} \subset \mathfrak{e}_8$.

\section{Resolution of Orbifold Singularities} \label{APP:resolution}

As stated in the text,
any chain of rational curves with self-intersections $-x_1$, \dots,
$-x_r$ ($x_j\ge2$)
is the resolution of $\mathbb{C}^2/\Gamma$, where $\Gamma=A(x_1,\dots,x_r)$
is the cyclic group of order $p$ generated by $(z_1,z_2)\mapsto (e^{2\pi i/p}z_1,
e^{2\pi iq/p}z_2)$ and the continued fraction expansion
of $p/q$ has coefficients $x_1$, \dots, $x_r$.  This goes back to
work of Jung \cite{jung} and Hirzebruch \cite{MR0062842}, with the resolution
described in modern language in \cite{Riemen:dvq}.

\begin{figure}[ptb]
\begin{center}
\includegraphics{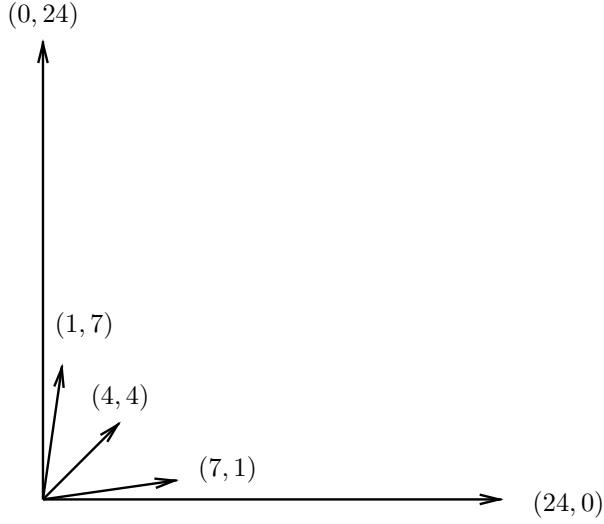}
\end{center}
\caption{Toric data for the $(4,2,4)$ model, with $p/q= 24/7$.}
\label{toricfigure}
\end{figure}

We illustrate with the example $p/q=24/7=4-1/(2-1/4)$, whose resolution we describe
in terms of toric geometry, as illustrated in figure~\ref{toricfigure}.
The toric data begins with vectors $v_0=(0,p)$ and $v_1=(q,1)$, and then
continues with $v_{j-1}+v_{j+1}=x_jv_j$ where $x_j$ are the continued
fraction coefficients, ending with $v_{r+1}=(p,0)$.
(In our example, this produces the five vectors shown in the figure.)
Each pair of adjacent vectors determines
a coordinate chart in which the Laurent monomial $z_1^mz_2^n$ extends to
a holomorphic function exactly if
$(m,n)\cdot v_{j-1}\ge0$ and $(m,n)\cdot v_{j}\le0$.  In our example, we get four coordinate charts in this way,
described as follows:
\begin{equation}
\begin{aligned}
 (\alpha_1,\beta_1)&=(z_1^{24},z_1^{-7}z_2)\\
 (\alpha_2,\beta_2)&=(z_1^7z_2^{-1},z_1^{-4}z_2^4)=(\beta_1^{-1},\alpha_1\beta_1^4)\\
 (\alpha_3,\beta_3)&=(z_1^4z_2^{-4},z_1^{-1}z_2^7)=(\beta_2^{-1},\alpha_2\beta_2^2)\\
(\alpha_4,\beta_4)&=(z_1z_2^{-7},z_2^{24})=(\beta_3^{-1},\alpha_3\beta_3^4)\\
\end{aligned}
\end{equation}
In general, there will be $r+1$ such coordinate charts $(\alpha_i,\beta_i)$.
The $i^{\text{th}}$ exceptional divisor appears in both the $i^{\text{th}}$
and the $(i+1)^{\text{st}}$ charts.

For the D-type orbifold group singularities, and their resolution, see \cite{Riemen:dvq, UTWOorb, Brieskorn}.

\newpage

\bibliographystyle{utphys}
\bibliography{onezero}

\end{document}